\documentclass[preprint,showpacs,preprintnumbers,amssymb,superscriptaddress,aps,prd,nofootinbib,11pt]{revtex4-1}
\usepackage{graphicx, epsfig, amssymb} 
\usepackage{amsmath, amsfonts}
\usepackage{bm} 
\usepackage{enumitem}
\usepackage[usenames]{color}
\definecolor{navyblue}{rgb}{0.0, 0.0, 0.5}
\usepackage[linktocpage,colorlinks=true,allcolors=navyblue]{hyperref}
\usepackage{subfigure}
\usepackage[utf8]{inputenc}
\usepackage{tensor}
\usepackage{soul}
\usepackage{pict2e}
\usepackage{diagbox}
\usepackage{empheq}

\usepackage{float}
\usepackage{comment}
\usepackage{array}
\usepackage{units}
\usepackage{wasysym}

\DeclareMathAlphabet{\mathpzc}{OT1}{pzc}{m}{it}

\newcommand{\nn}{\nonumber}
\newcommand{\beq}{\begin{equation}}
\newcommand{\eeq}{\end{equation}}
\newcommand{\qq}{\mathfrak{q}}

\newcommand{\Op}{\hat{\mathcal{D}}_2}
\newcommand{\Vg}{V_{(g)}}

\begin{document}

\title{Quasinormal modes of massive vector fields on the Kerr spacetime}
 
\author{Jake Percival}\email{jnpercival1@sheffield.ac.uk}
\affiliation{Consortium for Fundamental Physics,  School of Mathematics and Statistics,
University of Sheffield, Hicks Building, Hounsfield Road, Sheffield S3 7RH, United Kingdom \looseness=-1}
\author{Sam R. Dolan}\email{s.dolan@sheffield.ac.uk}
\affiliation{Consortium for Fundamental Physics,  School of Mathematics and Statistics,
University of Sheffield, Hicks Building, Hounsfield Road, Sheffield S3 7RH, United Kingdom \looseness=-1}

\begin{abstract}
We study the spectrum of quasinormal mode frequencies for a Proca field on a rotating black hole spacetime. First, we review how the introduction of field mass modifies the spectrum in the scalar-field case, leading to evanescent modes and to quasiresonance. Next, we examine the three physical polarizations of the Proca field and their relation to the electromagnetic field modes in the massless limit. Exploiting a separation of variables, we obtain a five-term recurrence relation from an appropriate ansatz for the radial function. Gaussian elimination and the modified Lentz algorithm are applied, and the quasinormal frequencies are computed from the roots of a continued fraction. We validate our method by calculating quasibound state frequencies, which are complementary to quasinormal modes, and which can be calculated using the same method. We present a selection of results for the low-lying overtones of all three polarizations, across a range of black hole spins and field masses.
\end{abstract}

\date{\today}

\maketitle

\section{Introduction}

A bell struck by a hammer vibrates in a characteristic manner, emitting a series of harmonics (or \emph{partials}) with a spectrum of frequencies and decay rates that are intrinsic to the bell itself, rather than to the hammer. Similarly, a perturbed black hole will return to a quiescent state by radiating through its natural damped resonances, known as quasinormal modes (QNMs) \cite{Vishveshwara:1970, Chandrasekhar:1998, Leaver:1985ax, Kokkotas:1999bd, Berti:2009kk, Konoplya:2011qq}. The aim of \emph{black hole spectroscopy} is to characterise such modes theoretically, and to extract QNM parameters from experimental data \cite{Berti:2005ys, Baibhav:2017jhs}. In fact, gravitational wave (GW) chirps from binary mergers, such as GW150914 \cite{Abbott:2016blz}, show the clear imprint of the $\ell = m = 2$ fundamental mode at late times in the ringdown phase. If two or more modes are identified \cite{Berti:2005ys, Giesler:2019uxc}, QNMs can be used to constrain the mass and spin of the black hole, and to test the no-hair theorem \cite{Isi:2019aib} and the general theory of relativity (GR) itself.

The QNM spectrum is the infinite set of complex frequencies $\omega_{\lambda} = \varpi_{\lambda} - i \Gamma_{\lambda}$, with $\varpi_{\lambda}$ the oscillation frequency and $\Gamma_{\lambda}$ the damping rate. Each mode in the spectrum is specified by a set of discrete numbers $\lambda$, where typically $\lambda = \{\ell, m, n, \mathcal{P}\}$, with $\ell$ and $m$ the angular momentum numbers, $n$ the overtone number and $\mathcal{P}$ the polarization state. The spectrum itself depends on the black hole parameters, such as the mass $M$ and angular momentum $J = a M$, and the properties of the perturbing field, such as its spin $s$ and its mass $\mu$. A key dimensionless parameter is
$
\frac{M \mu}{m_P^2} ,
$
which is of the same order as the ratio of the horizon radius of the black hole to the Compton wavelength of the field. 
Here $m_P$ is the Planck mass; henceforth we adopt units such that $G = c = m_P = 1$. 

The gravitational QNMs ($s=2$, $\mu=0$) of the Kerr black hole have been well-studied since the 1970s, due to their key role in the ringdown phase of black hole mergers. More widely, the study of QNMs of massless fields ($\mu = 0$) on a variety of black hole spacetimes has generated a substantial literature; see Refs.~\cite{Kokkotas:1999bd, Berti:2009kk, Konoplya:2011qq} for review articles. By comparison, the QNM spectrum of massive fields ($\mu \neq 0$) has received less attention. 

In 1991, Simone and Will \cite{Simone:1991wn} applied the WKB method to find quasinormal frequencies of the scalar field on the Schwarzschild and Kerr black hole spacetimes, finding that the scalar field mass $\mu$ led to an increase in oscillation frequency $\varpi$, and a decrease in damping $\Gamma$. Konoplya and Zhidenko \cite{Konoplya:2004wg, Konoplya:2005hr, Konoplya:2006br} found that, for large masses, the fundamental mode of the scalar field approaches a vanishing damping rate ($\Gamma \rightarrow 0$) and beyond a critical value the fundamental mode disappears from the spectrum. This phenomenon is known as \emph{quasiresonance} \cite{Ohashi:2004wr}. 

The QNMs of massive scalar fields have attracted further interest recently, motivated by the observation that a subfamily of Horndeski theories give rise to a scalar QNM spectrum characterized by a single parameter that acts as an effective mass \cite{Tattersall:2018nve, Lagos:2020oek}. In Ref.~\cite{Tattersall:2018nve} a series expansion of QNM frequencies in inverse powers of $L \equiv \ell + 1/2$ was obtained, extending the method of Ref.~\cite{Dolan:2009nk}.

The QNMs of massive fields of higher spin ($s > 0$) on Schwarzschild spacetime have also been studied. Cho \cite{Cho:2003qe} applied the WKB method to study the QNMs of the Dirac field ($s=1/2$) concluding,  as in the $s=0$ case, that $\varpi$ increases and $|\Gamma|$ decreases with the field mass $\mu$ (see also Ref.~\cite{Chang:2007aa} for the Reissner-Nordstr\"om spacetime). Konoplya \cite{Konoplya:2005hr} investigated the QNMs of the monopole mode of the Proca field. Rosa \& Dolan \cite{Rosa:2011my} studied the higher multipoles, identifying the three polarization states expected for a massive vector field. Brito, Cardoso and Pani \cite{Brito:2013wya} studied the massive spin-2 field on a black hole spacetime, and calculated QNM frequencies for the odd-parity axial sector (see Fig 2 in Ref.~\cite{Brito:2013wya}).

The QNMs of massive fields of higher spin ($s > 0$) on Kerr spacetime is a relatively unexplored arena. 
In Ref.~\cite{Konoplya:2017tvu}, the first numerical results for the QNMs of the massive Dirac field ($s=1/2$) on the Kerr spacetime were obtained via the Frobenius method. The introduction of mass $\mu \neq 0$ splits a degeneracy in the spectrum, leading to two polarizations with distinct QNM frequencies. 

To date, the QNMs of the Proca field ($s=1$, $\mu > 0$) on Kerr spacetime have not been calculated. The purpose of this paper is to fill this lacuna by exploiting the separation of variables for the Proca field on Kerr recently achieved by Frolov \emph{et al.} \cite{Frolov:2018ezx}\footnote{The modes labelled `quasinormal' in Ref.~\cite{Frolov:2018ezx} are the quasibound states in our nomenclature; see Sec.~\ref{sec:review}.}, who built on work by Lunin \cite{Lunin:2017drx}. The separability has already been used to the calculate of the spectrum of \emph{quasibound states} of the Proca field on Kerr \cite{Frolov:2018ezx, Dolan:2018dqv, Siemonsen:2019ebd, Baumann:2019eav} (see Refs.~\cite{Witek:2012tr, Pani:2012bp, Pani:2012vp, Baryakhtar:2017ngi,East:2017mrj, East:2017ovw, Cardoso:2018tly} for complementary approaches). In this work we show that both the quasinormal and quasibound spectra can be calculated by solving a particular five-term recurrence relation.

In Sec.~\ref{sec:review} we review the QNMs of the massive scalar field ($s=0$), in the Schwarzschild (\ref{sec:scalar-schw}) and Kerr (\ref{sec:scalar-kerr}) cases. Here we examine the association between QNMs and unstable circular orbits of geodesics, via the WKB method; and we distinguish propagative and evanescent modes. In Sec.~\ref{sec:proca}, we detail our method for calculating QNMs of the Proca field, covering the Kerr spacetime (\ref{sec:spacetime}), the method of separation of variables (\ref{sec:sep}), a new five-term recurrence relation for QNMs (\ref{sec:5term}); polarization states and angular eigenvalues (\ref{sec:polarization}); and the numerical methods used and their validation (\ref{sec:numerical}). Section \ref{sec:results} covers the main results, and we conclude with a discussion in Sec.~\ref{sec:conclusions}.

\section{Review: massive scalar QNMs\label{sec:review}}
In this section we review the QNMs of the massive scalar field on a static black hole spacetime, principally to develop an understanding of the effect of field mass $\mu$ on the spectrum in a base case, to set a foundation for an exploration of the Proca field on a spinning black hole spacetime. It is also relevant in light of recent work on QNMs in Horndeski gravity \cite{Lagos:2020oek,Tattersall:2018nve}.

\subsection{Scalar QNMs on Schwarzschild\label{sec:scalar-schw}}
The massive Klein-Gordon equation $(\Box - \mu^2) \Phi = 0$ on Schwarzschild spacetime is amenable to a separation of variables, $\Phi = r^{-1} u_{\ell \omega}(r) e^{-i \omega t} Y_{\ell m}(\theta,\phi)$, leading to the radial equation
\beq
\frac{d^2 u_{\ell \omega}}{d r_\ast^2} + \left\{ \omega^2 - V_{\ell}(r) \right\} u_{\ell \omega} = 0, \quad \quad V_{\ell}(r) = f \left( \mu^2 + \frac{\ell (\ell + 1)}{r^2} + \frac{2 M \beta}{r^2} \right) , \label{eq:scalar}
\eeq
where $V_{\ell}(r)$ is the effective potential, $r_\ast$  the tortoise coordinate defined by $dr_\ast / dr = f^{-1}$ with $f \equiv 1 - 2M/r$, and $\beta = 1$ for the scalar field. The IN mode satisfies the boundary condition
\beq
u_{\ell \omega}(r) \sim \begin{cases}
 e^{- i \omega r_\ast} ,  & r_\ast \rightarrow -\infty , \\
 A^{-}_{\ell \omega} e^{-i p r} r^{-i \chi} + A^{+}_{\ell \omega} e^{i p r} r^{i \chi}  , & r_\ast \rightarrow \infty ,
\end{cases}
\eeq
where $p \equiv \sqrt{\omega^2 - \mu^2}$ with $\text{Re}(p) > 0$ and $A^\pm_{l \omega}$ are complex coefficients.  
A quasinormal mode frequency $\omega_{\ell n}$ is such that $A^{-}_{\ell \omega_{\ell n}} / A^{+}_{\ell \omega_{\ell n}} = 0$. In other words, the mode is purely ingoing at the future horizon, and purely outgoing at future infinity.

The QNM spectrum may be calculated numerically by finding the roots of a certain continued-fraction equation, as detailed in Ref.~\cite{Dolan:2007mj}. The structure of the spectrum near the real axis may be understood by application of the WKB method \cite{Schutz:1985km, Iyer:1986np}. Schutz and Will \cite{Schutz:1985km} showed that, at lowest WKB order, the square of the QNM frequency is approximately
\beq
\omega^2_{\ell n} \approx V_{\ell}(r_0) - i (n+1/2) \sqrt{-2 V_{\ell}''(r_0)}  . \label{eq:WKB}
\eeq
Here $r_0$ is the radius of the peak of the effective potential barrier, where $V_{\ell}^\prime(r_0) = 0$ and $V_{\ell}^{\prime \prime} (r_0) < 0$. In other words, the low-$n$ QNMs are approximately determined by the height of the potential barrier and its second derivative only; and so where $V^{\prime \prime}_{\ell} \rightarrow 0$ a quasiresonance \cite{Ohashi:2004wr, Konoplya:2011qq} is anticipated.

In the eikonal regime ($\ell + 1/2 \gg 1$), there is an association between the effective potential $V_{\ell}(r)$ for the scalar field in Eq.~(\ref{eq:scalar}), and the effective potential $\Vg(r)$ for a null ($\mu = 0$) or timelike ($\mu > 0$) geodesic, viz.
\beq
\dot{r}^2 = E^2 - \Vg (r) , \quad \quad \Vg = f \left( \mu^2 + \frac{L^2}{r^2} \right) ,  \label{eq:geoV}
\eeq
where $E = -u_t$, $L = u_\phi$ and $u^\alpha = dx^\alpha / d\lambda$ is the geodesic tangent vector such that $g_{\alpha \beta} u^\alpha u^\beta = - \mu^2$. With the associations $E  \leftrightarrow \omega$ and $L \leftrightarrow \ell + 1/2$, the potentials $V_{\ell}(r)$ and $\Vg(r)$ match, up to terms involving neither $\mu$ nor $\ell + 1/2$. (Alternatively, setting $\beta = 0$ and making the usual Langer replacement $\ell (\ell + 1) \rightarrow (\ell + 1/2)^2$). 

The maximum (minimum) of the geodesic potential is associated with an unstable (stable) circular geodesic orbit. In turn, the unstable (stable) circular orbit is associated with the low overtones of the quasinormal mode (quasibound state) spectrum, via the correspondence above and the WKB formula (\ref{eq:WKB}). We shall now distinguish between \emph{propagative} modes with $\text{Re}(\omega^2) - \mu^2 > 0$ and \emph{evanescent} modes with $\text{Re}(\omega^2) - \mu^2 < 0$. In the geodesic picture, if the peak of the potential $\Vg(r_0)$ exceeds $\mu^2$ the mode is propagative; otherwise it is evanescent.

\begin{figure}
  \subfigure[\; Massless case]{\includegraphics[width=8cm]{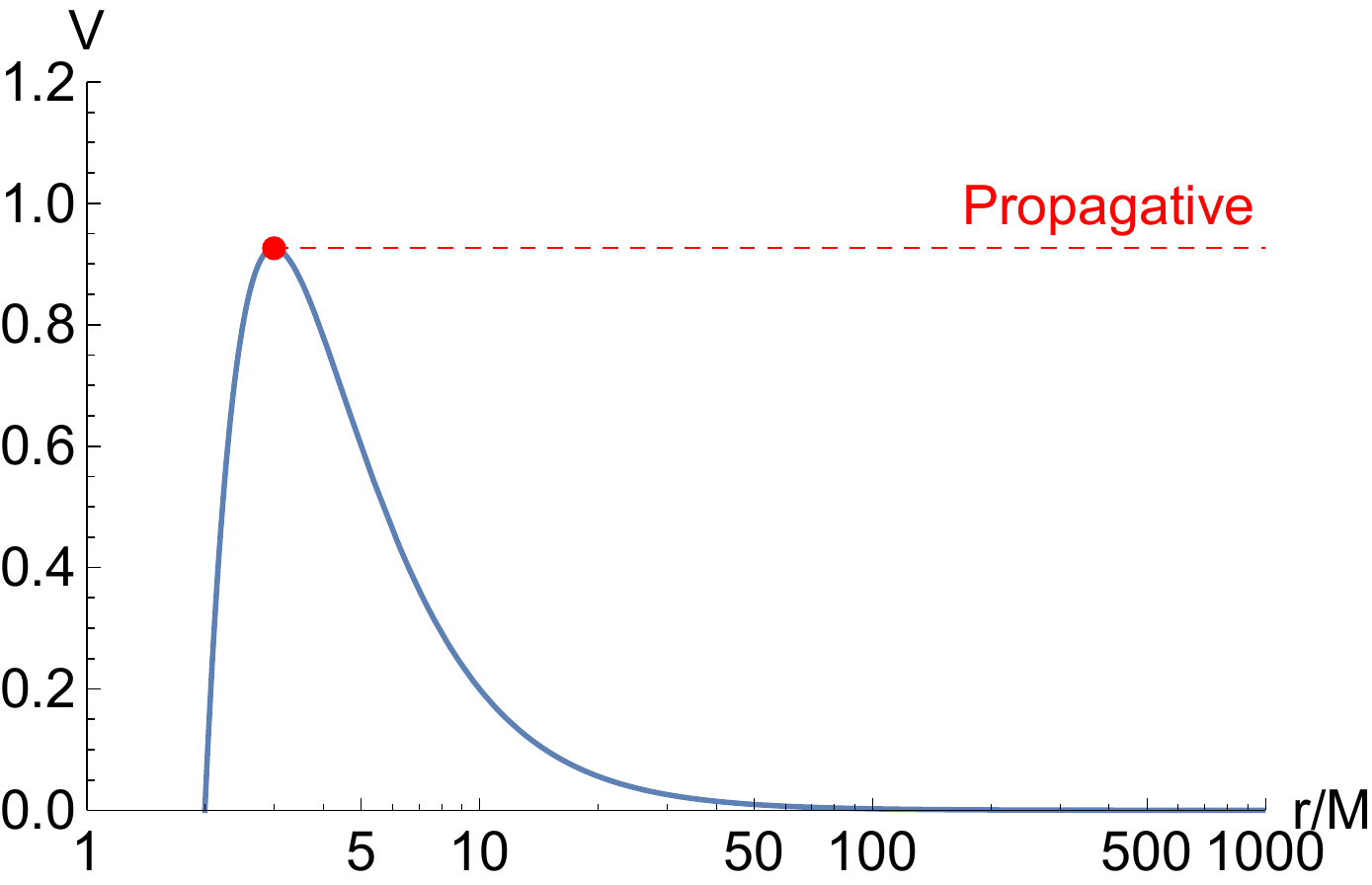}\label{fig:Vplot0}}
  \subfigure[\; Low-mass case]{\includegraphics[width=8cm]{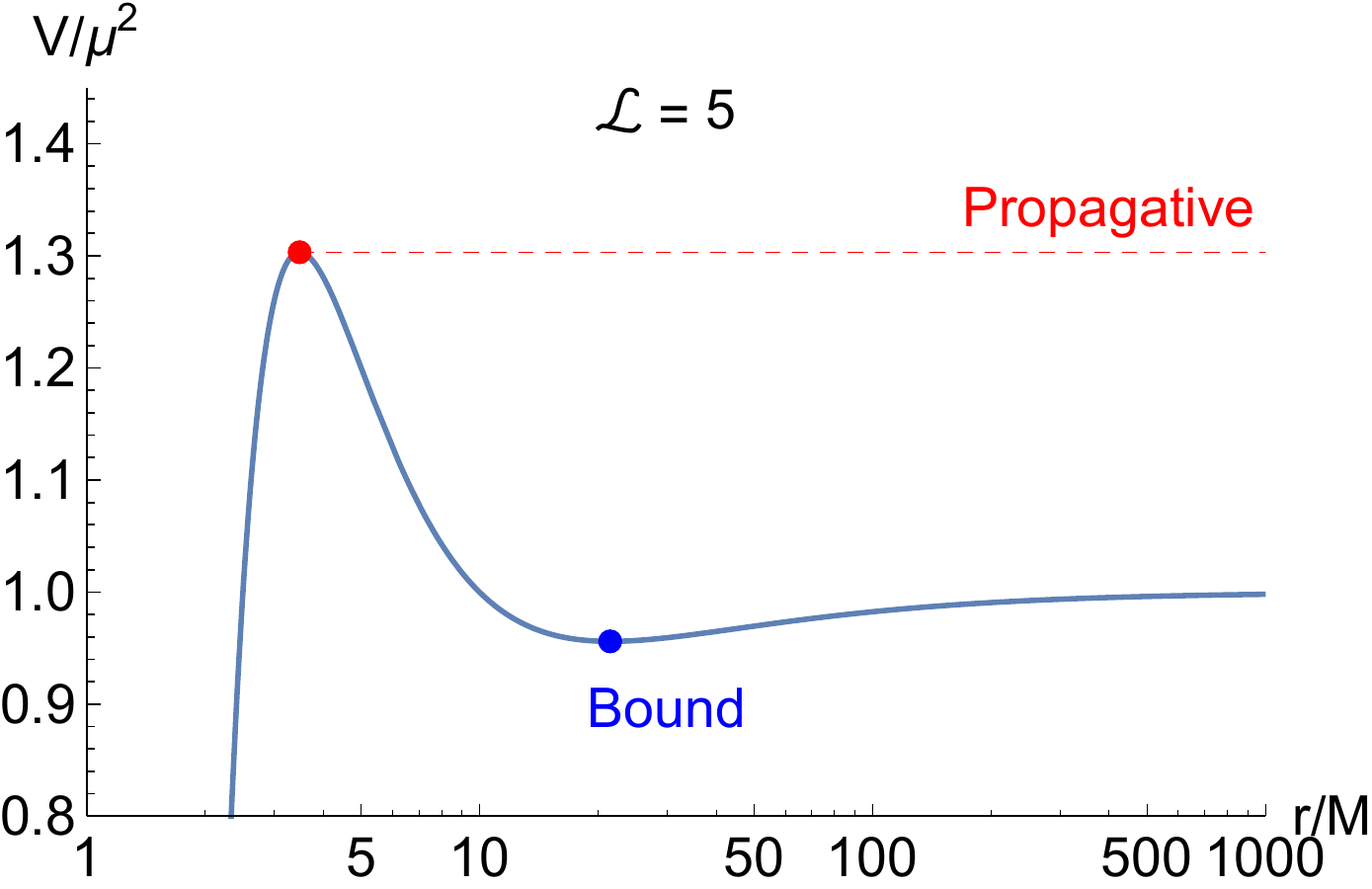}\label{fig:Vplot1}}
  \subfigure[\; Marginally-bound zoom-whirl (MBZW) case]{\includegraphics[width=8cm]{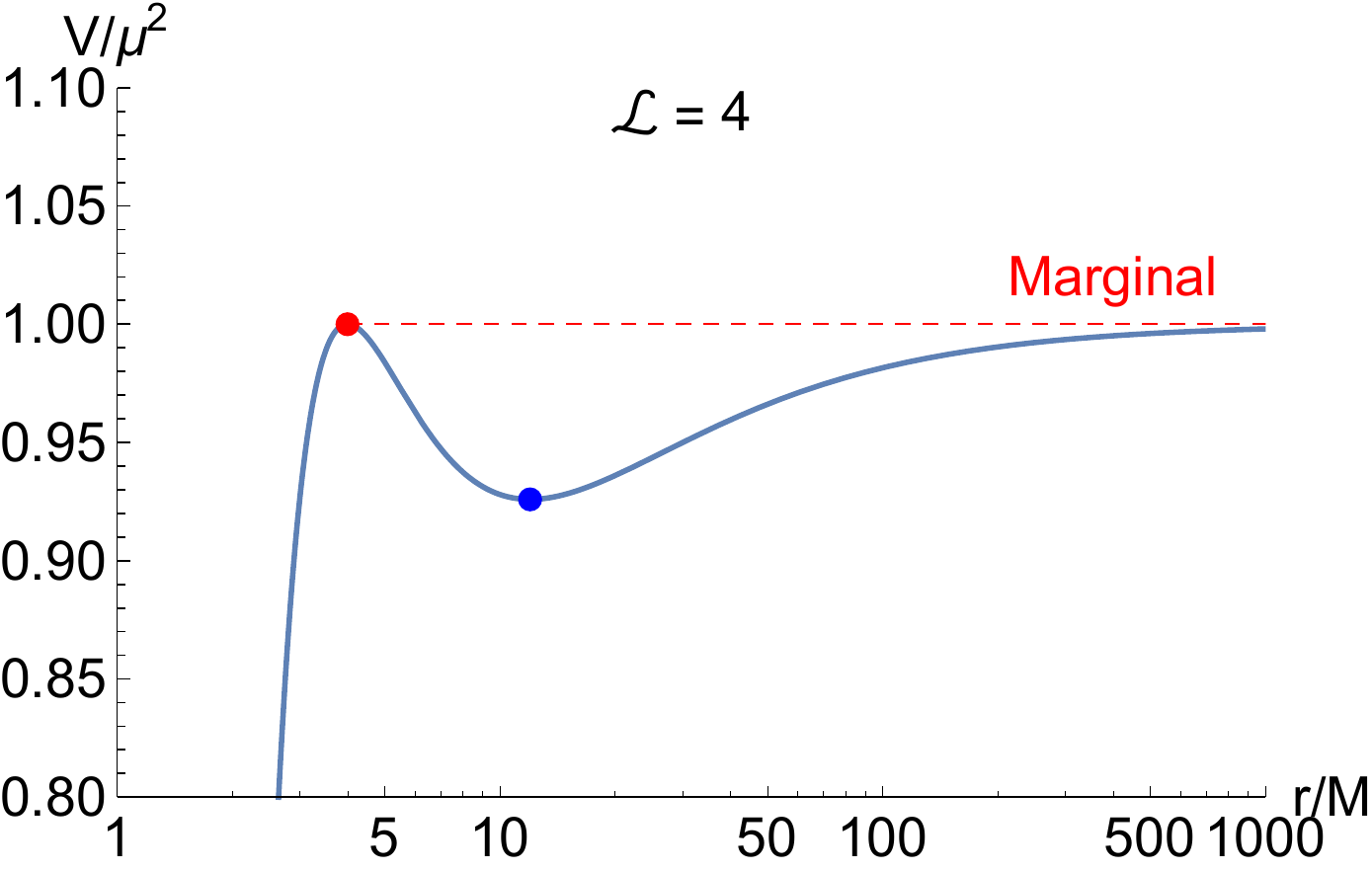}\label{fig:Vplot2}}
  \subfigure[\; Innermost stable circular orbit (ISCO) case]{\includegraphics[width=8cm]{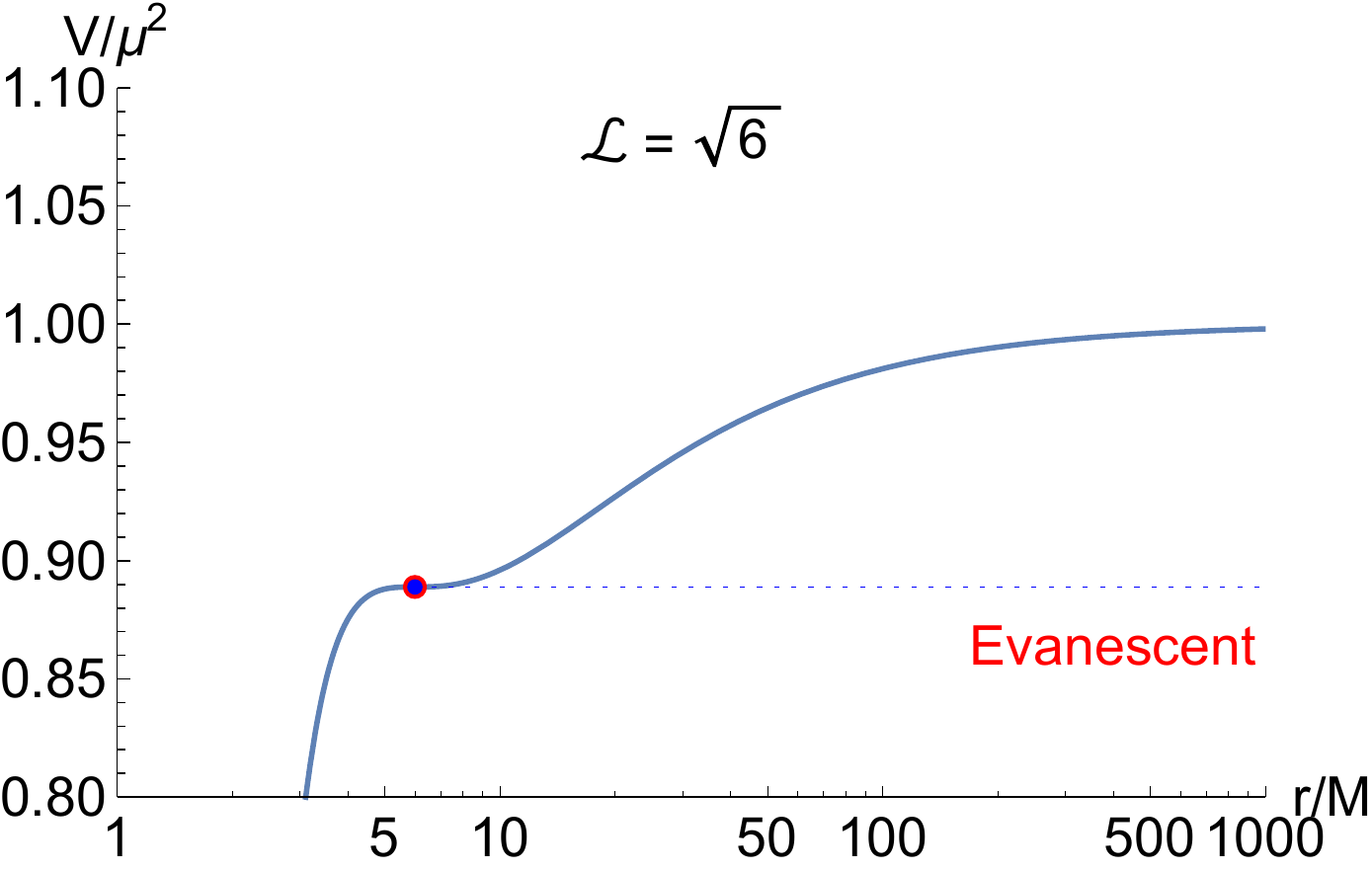}\label{fig:Vplot3}}
 \caption{The geodesic potential for several values of $\mathcal{L} = (\ell + 1/2) / M \mu$, showing the unstable and stable circular orbits (red and blue points) associated with quasinormal modes and quasibound states, respectively.}
 \label{fig:Vplot}
\end{figure}

Figure \ref{fig:Vplot} shows the geodesic potential $\Vg(r)$ for four values of $L / (M\mu)$. In the massless case (see Fig.~\ref{fig:Vplot0}), the maximum of $\Vg$ is at the photon orbit at $r_0 = 3M$. There is no quasibound spectrum in this case, due to the absence of a potential minimum. For a `small' mass (Fig.~\ref{fig:Vplot1}), there is a spectrum of propagative QNMs associated with the maximum (unstable circular orbits), and quasibound states associated with the minimum (stable circular orbits). Fig.~\ref{fig:Vplot2} shows the marginal case in which $\Vg(r_0) = \mu^2$ that separates propagative and evanescent QNMs. The associated geodesic is the marginally-bound zoom-whirl orbit. 
Fig.~\ref{fig:Vplot3} shows the case in which the stationary points come together to form an inflexion. At lowest WKB order the QNM frequency is real (as $V''(r_0) = 0$), corresponding to a quasiresonance. This QNM mode is evanescent, and the associated geodesic is the innermost stable circular orbit (ISCO). 

\begin{figure}
  \includegraphics[width=15cm]{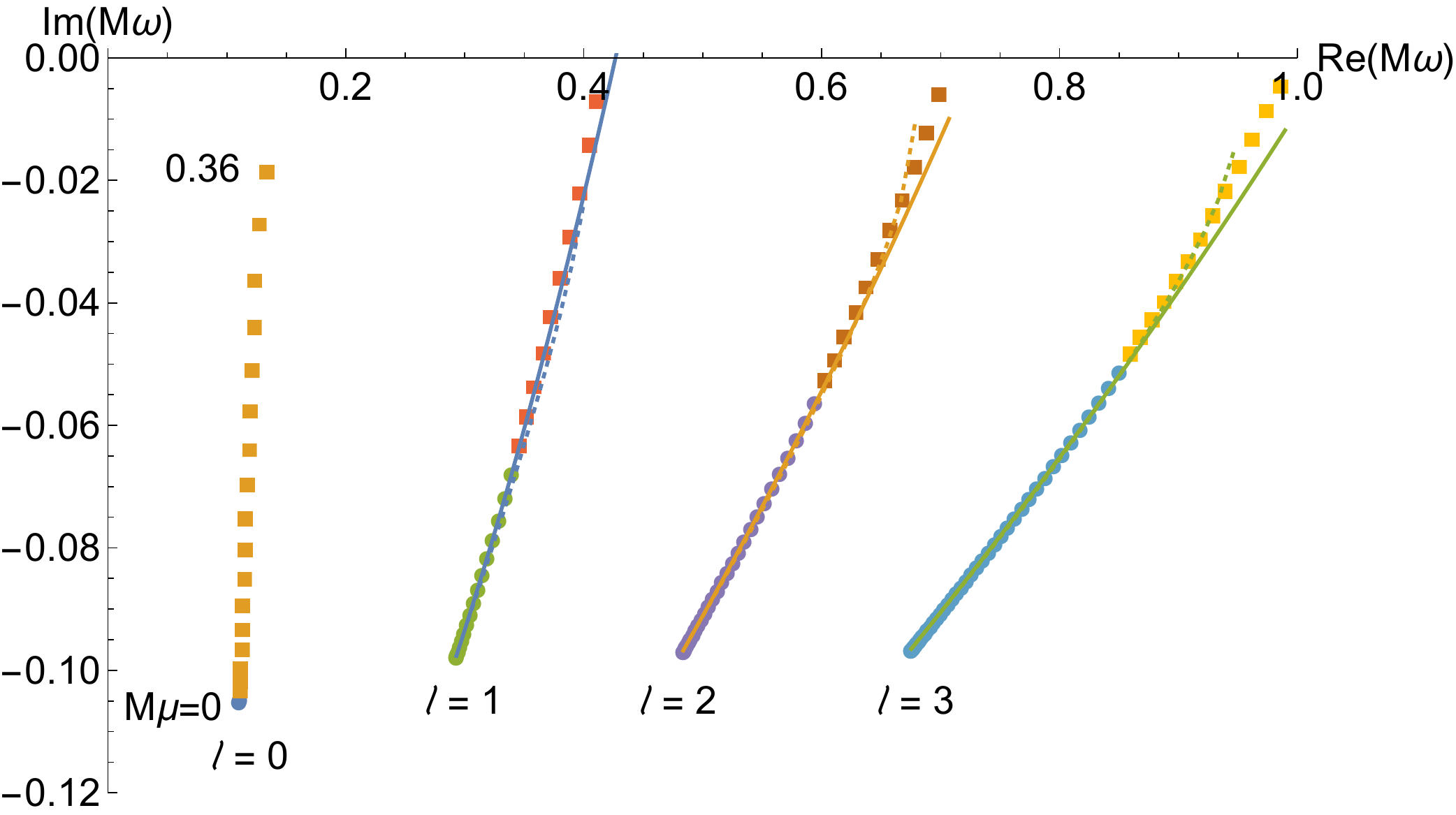}
  \caption{
  Fundamental ($n=0$) quasinormal mode frequencies of the massive scalar field ($s=0$) on Schwarzschild spacetime. Circles (squares) indicate propagative (evanescent) modes calculated with the continued-fraction method \cite{Dolan:2007mj}, for masses $M\mu = 0.02 k$ ($k \in \mathbb{N}$). The last mode shown is at $M\mu = 0.36$, $0.52$, $0.8$ and $1.1$, for multipoles $\ell = 0,1,2$ and $3$ respectively. The solid line shows the approximation of Eq.~(29) in Ref.~\cite{Tattersall:2018nve}; the dashed line the leading-order WKB approximation of Eq.~(\ref{eq:WKB}) using the geodesic potential (\ref{eq:geoV}). 
  }
  \label{fig:s0QNMschw}
\end{figure}

Figure \ref{fig:s0QNMschw} shows numerically-determined fundamental ($n=0$) QNM frequencies for the massive scalar field on Schwarzschild. As the mass $\mu$ is increased, the modes increase in frequency and move towards the real axis. The transition from propagative to evanescent is indicated by a change of symbol. At a critical value of $M\mu$, close to that associated with the ISCO, the branch of QNMs disappears. 

The lowest-order WKB approximation does well in describing the migration of the fundamental frequencies in the complex-$\omega$ plane for $\ell \gtrsim 2$. The WKB frequencies are shown in Fig.~\ref{fig:s0QNMschw} by a dashed line (N.B.~here we have used the geodesic potential $\Vg(r)$ in place of the scalar field potential $V_\ell(r)$ in Eq.~(\ref{eq:WKB})). A higher-order approximation, obtained in Ref.~\cite{Tattersall:2018nve} and based on the method of Ref.~\cite{Dolan:2009nk}, is shown as a solid line.

The response of a black hole to a wave-packet of a massive scalar field was investigated in Refs.~\cite{Decanini:2014bwa, Decanini:2016ifm}. In the low-mass regime ($M\mu \ll \ell$), QNM ringing can be clearly identified, as in the massless case. However, outside this regime the quasibound states and an oscillatory power-law tail also play a role, and an unambiguous identification of the different contributions in the response is not straightforward. Although Ref.~\cite{Decanini:2014bwa} identified `giant ringings' for certain QNMs, in Ref.~\cite{Decanini:2016ifm} it was established that giant ringings are not significant in practice, as they arise in evanescent modes at late times, and cannot be easily separated from the other contributions.

\subsection{Scalar QNMs on Kerr\label{sec:scalar-kerr}}
The QNM frequencies of the massive scalar field on Kerr were calculated approximately via a WKB expansion in Ref.~\cite{Simone:1991wn}, and more precisely via a three-term recurrence relation in Ref.~\cite{Konoplya:2006br}; and later also in Ref.~\cite{Dolan:2007mj}. The geometrical intepretation of the massless spectrum in the eikonal limit was explored in \cite{Dolan:2010wr,Yang:2012he,Yang:2012pj,Yang:2013uba}.

\begin{figure}
  \includegraphics[width=15cm]{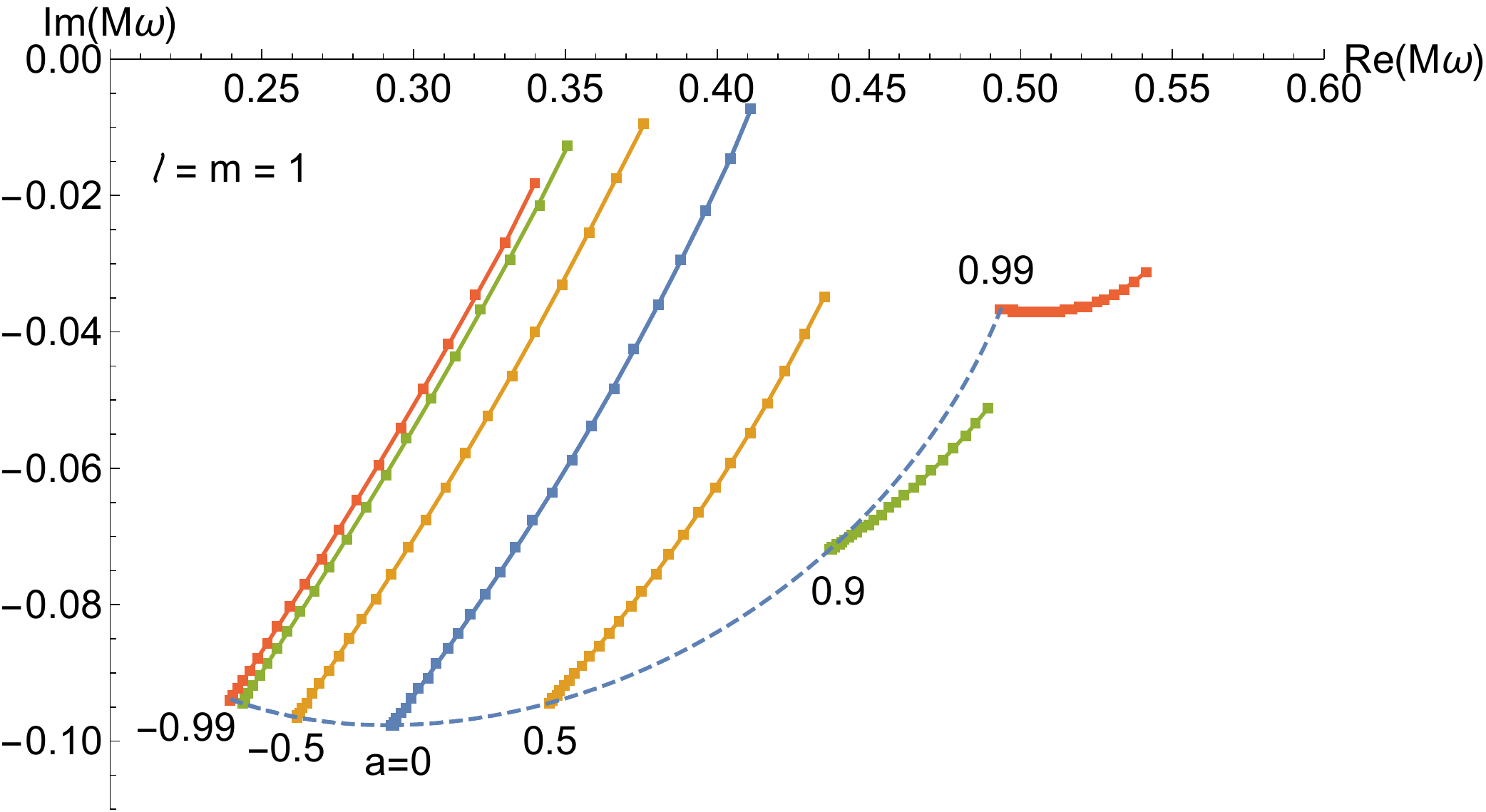}
  \caption{
  Fundamental ($n=0$) quasinormal mode frequencies of the massive scalar field ($s=0$) on Kerr spacetime for $\ell = m = 1$ and spin parameters $a/M \in \{-0.99,-0.9,-0.5,0,0.5,0.9,0.99\}$. The dashed line show the massless spectrum. The points are for masses $M\mu = 0.02 k$ ($k \in \mathbb{N}$) up to $\{ 0.40, 0.42, 0.46, 0.52, 0.52, 0.52, 0.64 \}$ reading left to right.
  }
  \label{fig:s0QNMkerr}
\end{figure}

Figure \ref{fig:s0QNMkerr} shows the spectrum of the modes $\ell = m = 1$. The rotation of the black hole $a$ splits the degeneracy on $m$, the azimuthal number. As in the $a=0$ case, the field mass $\mu$ typically leads to an increase in oscillation frequency and a decrease in damping, although the $a=0.99$, $m=1$ case shows a slight increase in damping for small $M \mu$ (see also Fig.~1 in Ref.~\cite{Dolan:2007mj}). 

\section{Proca QNMs on Kerr spacetime\label{sec:proca}}

 \subsection{Kerr spacetime and the principal tensor\label{sec:spacetime}}
 
The Kerr spacetime in the Boyer-Lindquist coordinate system $\{t,r,\theta,\phi\}$ is described by the line element 
\begin{align}
ds^2 
 &= - \left(1-\frac{2Mr}{\Sigma}\right)dt^{2} - \frac{4aMr\sin^{2}\theta}{\Sigma}dtd\phi
 + \frac{\Sigma}{\Delta}dr^{2} + \Sigma d\theta^{2}
 + \left[\left(r^{2}+a^{2}\right)+\frac{2Mr}{\Sigma}a^{2}\sin^{2}\theta\right] \sin^{2}\theta \, d\phi^{2}
\end{align}
 where
\begin{align*}
 \Delta &\equiv r^{2}-2Mr+a^{2} = (r-r_+)(r-r_-) , & \Sigma &\equiv r^{2}+a^{2}\cos^{2}\theta ,
\end{align*}
and $r_{\pm} \equiv M \pm \sqrt{M^2 - a^2}$. 
The Kerr spacetime is stationary and axisymmetric; these explicit symmetries are represented by Killing vectors $\partial_{t}$ and $\partial_{\phi}$. It also has \emph{hidden} symmetries \cite{Frolov:2017kze}. In particular the Kerr spacetime belongs to the family of Kerr-NUT-(A)dS spacetimes and thus it admits a non-degenerate, closed, conformal, Killing-Yano 2-form $h_{a b}$ known as the \emph{principal tensor} \cite{Frolov:2017kze}, which is the Hodge dual of the Killing-Yano tensor $f_{a b}$. For Kerr spacetime,
\begin{subequations}
\begin{align}
g^{ab} &= \phantom{a \cos \theta} \frac{\Delta}{\Sigma} l^{(a}_{+} l^{b)}_{-} + \phantom{i a \cos \theta} \frac{1}{\Sigma} m^{(a}_{+} m^{b)}_{-} , \\
f^{ab} &= a \cos \theta  \frac{\Delta}{\Sigma}  l^{[a}_{+} l^{b]}_{-} + \quad \quad \; i r \frac{1}{\Sigma} m^{[a}_{+} m^{b]}_{-} , \\
h^{ab} &= \quad \; \;  -r \frac{\Delta}{\Sigma}  l^{[a}_{+} l^{b]}_{-} + i a \cos \theta \frac{1}{\Sigma} m^{[a}_{+} m^{b]}_{-} , 
\end{align}
\end{subequations}
where 
\begin{align}
l^a_\pm &\equiv \left[ \pm (r^2+a^2) / \Delta, 1, 0, \pm a / \Delta \right] , &
m^a_{\pm} &\equiv \left[ \pm i a \sin \theta, 0, 1, \pm i \csc \theta \right] .
\label{eq:tetrad2}
\end{align}
Here, round (rectangular) parantheses denote the symmetrized (anti-symmetrized) tensors. 
The principal tensor is used in the construction of the separable ansatz of the Proca equation in this spacetime.

 \subsection{Proca field: Separation of variables\label{sec:sep}}
 
 The Proca equation for a massive vector field $A^{a}(x)$ is
 \beq
 \nabla_{b}F^{a b}+\mu^{2}A^{a}=0 , \label{eq:Proca}
 \eeq
 where $F^{a b}$ is the Faraday tensor defined by
$
 F_{a b}=\nabla_{a}A_{b}-\nabla_{b}A_{a}.
$
A non-zero mass $\mu$ removes any ambiguity in the choice of gauge, as $\nabla_{a}A^{a}=0$ (the Lorenz gauge condition) is implied by taking the divergence of the field equations (\ref{eq:Proca}). 

The Lunin-Frolov-Krtou{\v{s}}-Kubiz{\v{n}}\'ak (LFKK) ansatz \cite{Frolov:2018ezx, Lunin:2017drx} used to separate the field equation in the Kerr spacetime is
\beq
A^{a}=B^{ab}\nabla_{b}Z,  \label{eq:ansatz}
\eeq
where $Z$ is a scalar field, and $B^{ab}$ is the polarization tensor defined by
\beq
B^{ab}(g_{bc}+i\nu h_{bc}) = \delta^{a}_{c} . \label{eq:Bdef}
\eeq
Here $\nu$ is a separation constant, henceforth referred to as the \emph{angular eigenvalue}. By solving Eq.~(\ref{eq:Bdef}), one obtains an explicit expression for the tensor $B^{ab}$ in Eq.~(\ref{eq:ansatz}) given by \cite{Frolov:2018pys, Dolan:2018dqv}
\beq
B^{ab} = \frac{\Delta_r}{2\Sigma} \left( \frac{l_+^a l_-^b }{1-i\nu r} + \frac{l_-^a l_+^b }{1+i\nu r} \right) + \frac{1}{2\Sigma} \left( \frac{m_+^a m_-^b }{1-\nu a \cos \theta} + \frac{m_-^a m_+^b }{1+\nu a \cos \theta} \right) .
\label{eq:Babalt}
\eeq

Frolov \emph{et al.}~\cite{Frolov:2018ezx} showed that the field equation admits separable solutions of the form $Z = R(r) S(\theta) \exp(-i \omega t + i m \phi)$, leading to second-order ODEs for the radial and angular functions,
\begin{subequations}
\begin{eqnarray}
\qq_r \frac{d}{dr} \left[ \frac{\Delta}{\qq_r} \frac{dR}{dr} \right] + \left[ \frac{K_r^2}{\Delta} + \frac{2-\qq_r}{\qq_r} \frac{\sigma}{\nu} - \frac{\qq_r \mu^2}{\nu^2} \right] R(r) &=& 0 ,  \label{eq:radial} \\
 \frac{\qq_\theta}{\sin \theta} \frac{d}{d\theta} \left[ \frac{\sin \theta}{\qq_\theta} \frac{dS}{d\theta} \right] - \left[ \frac{K_\theta^2}{\sin^2 \theta} + \frac{2-\qq_\theta}{\qq_\theta} \frac{\sigma}{\nu} - \frac{\qq_\theta \mu^2}{\nu^2} \right] S(\theta) &=& 0 ,  \label{eq:angular}
\end{eqnarray} 
\label{eq:FKKS}
\end{subequations}
where 
\begin{align}
K_r &= (a^2+r^2)\omega - am, & \qq_r &= 1+\nu^2 r^2,  \nn \\
K_\theta &= m - a\omega \sin^2 \theta, & \qq_\theta &= 1 - \nu^2 a^2 \cos^2 \theta, & \sigma &= \omega + a\nu^2 (m-a\omega). \label{eq:sigma}
\end{align}
In Ref.~\cite{Baumann:2019eav} the radial equation was cast into a form that highlights the existence of \emph{five} singular points in the complex plane, viz.,
\begin{align}
\frac{d^{2}R}{dr^{2}}+\left(\frac{1}{r-r_{+}}+\frac{1}{r-r_{-}}-\frac{1}{r-i/\nu}-\frac{1}{r+i/\nu}\right)\frac{dR}{dr} \nn \\
+\Biggl[-\frac{\Lambda}{\Delta}-q^{2}+\frac{\rho_{+}^{2}}{\left(r-r_{+}\right)^{2}}+\frac{\rho_{-}^{2}}{\left(r-r_{-}\right)^{2}}-\frac{A_{+}}{\left(r_{+}-r_{-}\right)\left(r-r_{+}\right)} \nn \\
+\frac{A_{-}}{\left(r_{+}-r_{-}\right)\left(r-r_{-}\right)}-\frac{\sigma}{\nu}\frac{r}{\Delta\left(r-i/\nu\right)}-\frac{\sigma}{\nu}\frac{r}{\Delta\left(r+i/\nu\right)}\Biggr]R & =0 , \label{eq:odeR}
\end{align}
where
\begin{align}
\Lambda &= \frac{\mu^{2}}{\nu^{2}}-\frac{\sigma}{\nu}+2a\omega m-a^{2}\omega^{2} ,  \\
A_{\pm} &= \rho_{+}^{2}+\rho_{-}^{2}+\frac{1}{4}\left(r_{+}-r_{-}\right)^{2}\left(\mu^{2}-\omega^{2}\right)+\left[M^{2}\left(\mu^{2}-7\omega^{2}\right)\pm M \left(r_{+}-r_{-}\right)\left(\mu^{2}-2\omega^{2}\right)\right] \label{eq:Lambda} \\
\rho_{\pm} &= \frac{2Mr_{\pm}\omega-am}{r_{+}-r_{-}}.  \label{eq:rhopm}
\end{align}
Regular singular points are located on the real axis at $r_{+},r_{-}$ and in the complex plane at $r=\pm i/\nu$. There is a confluent singular point at $r=\infty$. 

In Ref.~\cite{Dolan:2018dqv}, the angular equation was rewritten in the form
\begin{align}
\left(1-a^2 \nu^{2} \cos^2 \theta \right)\left[ \frac{d^{2}}{d\theta^{2}}+\cot \theta \frac{d}{d\theta} -\frac{m^{2}}{\sin^{2}\theta}+\Lambda\right] S \quad \quad & \nn \\
+ \left\{ q^{2}a^{4} \nu^{2} \cos^4 \theta - \left(q^{2}+2\sigma\nu\right)a^{2} \cos^{2}\theta - 2 a^{2} \nu^{2}\sin \theta \cos \theta \frac{d}{d\theta} \right\} S &=0 \label{eq:S2}
\end{align}
where $q^{2} \equiv \mu^{2}-\omega^{2}$.

 \subsection{Five-term recurrence relation\label{sec:5term}}
In this section we show that the problem of finding QNMs and bound states of the Proca field is equivalent to that of finding convergent solutions of five-term recurrence relations,
\beq
\alpha_n a_{n+2} + \beta_n a_{n+1} + \gamma_n a_n + \delta_n a_{n-1} + \epsilon_n a_{n-2} = 0, \quad \quad n \ge 2 , \label{eq:5term-expansion}
\eeq
where $a_n$ are series coefficients in the solution, and coefficients $\alpha_n, \ldots, \epsilon_n$ depend implicitly on the parameters including $\omega$ and $\nu$.


First we specify an \emph{ansatz} for the radial function $R(r)$ that respects the physical boundary conditions as $r \rightarrow r_+$ and $r \rightarrow \infty$. This is of the form
\beq
R\left(r\right)=\left(\frac{r-r_{+}}{r-r_{-}}\right)^{-i\rho}\left(r-r_{-}\right)^{\chi}e^{qr}\sum_{k=0}^{\infty}a_{k}\left(\frac{r-r_{+}}{r-r_{-}}\right)^{k}.
\eeq
where
\beq
q=\pm \sqrt{\mu^{2}-\omega^{2}}, \quad \quad 
\chi =\frac{M\left(2\omega^{2}-\mu^{2}\right)}{q}.  \label{eq:qdef}
\eeq
The parameter $q$ in the exponential depends on the boundary condition imposed far away from the black hole, with the choice $\text{Re}(q)>0$ for QNMs and $\text{Re}(q)<0$ for quasibound states. 

The parameter $\rho_+$ is determined by substituting the ansatz into the radial equation (\ref{eq:odeR}) and expanding around $r=r_{+}$. The requirement that the leading term in the radial equation vanishes yields $\rho=\pm \rho_{+}$, where $\rho_+$ is defined in Eq.~(\ref{eq:rhopm}). We demand that the field is regular on the future horizon in a coordinate system that is horizon-regular, such as ingoing-Kerr coordinates; this necessitates the choice $\rho=\rho_{+}$.

The recurrence relation (\ref{eq:5term-expansion}) for the coefficients $a_{k}$ is found by expanding the equation in powers of $x = \frac{r-r_+}{r-r_-}$ and solving it term by term. These manipulations were performed with the help of the symbolic algebra package Mathematica. 
For brevity we have defined $b=\sqrt{M^2-a^2}$, $u_{\pm}=1+\nu^{2}r_{\pm}^{2}$, $t_{\pm}=1\pm\nu^{2}r_{+}r_{-}$ and $c_{\pm}=1+Mr_{\pm}\nu^{2}$. The coefficients in the five-term relation, Eq.~(\ref{eq:5term-expansion}), are

{\small
\begin{subequations}
\begin{align}
\alpha_{n} & =-16b^{2}\left(n+2\right)q^{2}u_{+} \left(n+2-2i\rho_{+}\right) \\
\beta_{n} & =4bq \Biggl\{16b\left(n+1\right)^{2}qc_{+}+(A_{-} - A_{+}) u_{+}\left(1-2i\rho_{+}\right)+4bqA_{+}u_{+} \nn \\
& -2\left(n+1\right)\left[u_{+}\left(A_{+}-A_{-}\right)+8bq\left(b\left(q+r_{+}\left(qr_{+}-1\right)\nu^{2}\right)+2ic_{+}\rho_{+}\right)\right] \nn \\
& -4bq\left[-\Lambda+2bqu_{+}\left(1-2i\rho_{+}\right)-2ir_{-}r_{+}\nu^{2}\rho_{+}+2\rho_{+}^{2}+r_{+}^{2}\nu\left(-\Lambda\nu+2\nu\rho_{+}\left(i+\rho_{+}\right)-2\sigma\right)\right]\Biggr\}\\
\gamma_{n} & =- \Biggl\{ u_{+}\left(A_{-}-A_{+}\right)^{2}+8A_{-}bq\left(1+n\left(3+r_{+}\left(2M+r_{-}\right)\nu^{2}\right)-3i\rho_{+}+r_{+}\nu^{2}\left(r_{+}-i\left(2M+r_{-}\right)\rho_{+}\right)\right) \nn \\
& +8A_{+}bq\left(-1+4bqt_{+}+n\left(-3-r_{+}\left(2M+r_{-}\right)\nu^{2}\right)+3i\rho_{+}+ir_{+}\nu^{2}\left(2r_{-}\rho_{+}+r_{+}\left(i+\rho_{+}\right)\right)\right) \nn\\
& -16b^{2}q^{2}\Biggl[-2t_{+}\Lambda+n^{2}\left(-6-\left(4M^{2}+2r_{+}r_{-}\right)\nu^{2}\right)+8bn\left(-M\nu^{2}+qt_{+}\right)-u_{+}\rho_{-}^{2} \nn \\
& +2in\left(6+\left(4M^{2}+2r_{+}r_{-}\right)\nu^{2}\right)\rho_{+}+\rho_{+}\left(8ibM\nu^{2}-8ibqt_{+}+5\rho_{+}+r_{+}\left(2M+3r_{-}\right)\nu^{2}\rho_{+}\right)-4r_{-}r_{+}\nu\sigma\Biggr]\Biggr\}\\
\delta_{n} & =2 \Biggl\{ t_{+}\left(A_{-}-A_{+}\right)^{2}+32 b^2 \left(n-1\right)^{2}q^{2}c_{-}-2A_{-}bq\left(-1+6i\rho_{+}+r_{-}\nu^{2}\left(-2b-r_{+}+2i\left(2M+r_{+}\right)\rho_{+}\right)\right) \nn\\
& +2A_{+}bq\left(-1+4bqu_{-}+6i\rho_{+}+r_{-}\nu^{2}\left(-2b-r_{+}+2i\left(2M+r_{+}\right)\rho_{+}\right)\right) \nn\\
& -4b\left(n-1\right)q\left[A_{-}\left(-3-r_{-}\left(2M+r_{+}\right)\nu^{2}\right)+A_{+}\left(3+r_{-}\left(2M+r_{+}\right)\nu^{2}\right)+8bq\left(b\left(q+r_{-}\left(qr_{-}-1\right)\nu^{2}\right)+2ic_{-}\rho_{+}\right)\right] \nn \\
& -8b^{2}q^{2}\left[-\Lambda-2\rho_{-}^{2}-2bqu_{-}\left(1+2i\rho_{+}\right)+4\rho_{+}^{2}+2r_{-}r_{+}\nu^{2}\left(-\rho_{-}^{2}+\rho_{+}\left(i+\rho_{+}\right)\right)-r_{-}^{2}\nu\left(\Lambda\nu-2\nu\rho_{+}\left(\rho_{+}-i\right)+2\sigma\right)\right]\Biggr\}  \\
\epsilon_{n} & =-u_{-} \left(A_{-}-A_{+}+4bq\left(n-2+i\rho_{-}-i\rho_{+}\right)\right)\left(A_{-}-A_{+}+4bq\left(n-i\left(-2i+\rho_{-}+\rho_{+}\right)\right)\right)
\end{align}
\label{eq:5term}
\end{subequations} 
}

In Ref.~\cite{Leaver:1990zz}, Leaver conjectured that the smallest number of terms in a recurrence relation is related to the number of singular points in the differential equation. A five-term relation (\ref{eq:5term}) from a differential equation with 1 confluent and 4 regular singular points (\ref{eq:odeR}) is consistent with the conjecture.



 \subsection{Polarization states and the angular eigenvalue spectrum\label{sec:polarization}}
 
The Proca field has three distinct polarizations, in contrast to the two polarizations of the electromagnetic field. For given angular momentum numbers $\ell$, $m$ and overtone number $n$, there is  one odd-parity mode and two even-parity modes (see Ref.~\cite{Rosa:2011my} for the Schwarzschild case). The odd-parity mode and one of the even-parity modes are of `vector' type, and the remaining even-parity mode is of `scalar' type \cite{Rosa:2011my}. In the massless limit ($\mu \rightarrow 0$), the vector-type even-parity and odd-parity QNMs are degenerate (isospectral), matching with the $(\ell,m,n)$ QNM frequency of the electromagnetic field \cite{Rosa:2011my}. In the same limit, the scalar-type even-parity frequency matches with the corresponding $(\ell, m, n)$ QNM frequency of a massless scalar field ($s=0$).

The three polarizations correspond to three distinct values of the angular eigenvalue $\nu$, for each $(\ell, m, n)$. Below we make a closer inspection of the angular eigenvalue spectrum.

\subsubsection{The massless limit ($\mu \rightarrow 0$)}


The scalar-type mode is a pure-gauge mode in the massless limit, that is, $A_a =\nabla_a Z$ for some scalar function $Z$. Consequently, $F_{ab}=0$ and the Lorenz-gauge condition $\nabla_a A^a = 0$ implies that $Z$ satisfies the Klein-Gordon equation ($\nabla_a \nabla^a Z = 0$). By comparing $A_a =\nabla_a Z$ with ansatz (\ref{eq:ansatz}), we conclude that $B^{ab} = g^{ab}$ in this case, and hence it follows from Eq.~(\ref{eq:Bdef}) that $\nu \rightarrow 0$ for all scalar-type modes in the massless limit.

The angular eigenvalue $\nu$ for the vector modes can be found in the massless limit by appeal to the Teukolsky formalism \cite{Teukolsky:1973ha}. In Refs.~\cite{Dolan:2018dqv,Dolan:2019hcw} it was established that $\lambda$, the separation constant in the $s=-1$ Teukolsky equations, is related to $\nu$ by the following:
 \begin{align}
 \nu &  
 =  \frac{-2 \omega}{\lambda \mp \mathcal{B}}
 = \frac{\lambda \pm \mathcal{B}}{2 a (m - a \omega)} ,
 & \mathcal{B} 
 & =\sqrt{\lambda^{2}+4am\omega-4a^{2}\omega^{2}} ,
 \label{eq:massless-eig}
 \end{align}
 where $\mathcal{B}$ is the Teukolsky-Starobinski constant. As can be seen above, a single value of $\lambda$ generates two values of $\nu$. The lower (upper) sign gives the angular eigenvalue for the even-parity (odd-parity) vector mode. In the static limit ($a \rightarrow 0$), $\nu$ diverges for the odd-parity mode, and $\nu=\frac{-\omega}{\ell (\ell + 1)}$ for the even-parity mode.

\subsubsection{The Schwarzschild limit ($a = 0$)}
 In the static limit ($a \rightarrow 0$), the second term in Eq.~(\ref{eq:S2}) vanishes when $\nu$ is regular, and thus Eq.~(\ref{eq:S2}) reduces to the general Legendre equation. For angular functions that are regular at the poles, it follows that $\Lambda = \ell (\ell + 1)$, with $\ell \in \mathbb{N}$ and $\Lambda$ is defined in terms of $\nu$ in Eq.~(\ref{eq:Lambda}). Rearranging this yields two values of $\nu$, viz., 
\beq
\nu = \frac{-\omega}{\ell (\ell+1)}\frac{1\pm\sqrt{1+4 \ell (\ell +1) \mu^{2}/\omega^{2}}}{2} .  \label{eq:nustatic}
\eeq
The lower (upper) sign gives the angular eigenvalue for the even-parity scalar (even-parity vector) mode. In the massless limit, these reduce to $\nu=0$ and $\nu=\frac{-\omega}{\ell (\ell + 1)}$, respectively. 

The eigenvalue of the odd-parity mode diverges in the static limit. In the massless case, Eq.~(\ref{eq:massless-eig}) implies that in the static limit ($a \rightarrow 0$) the quantities $a\nu$, $\frac{\sigma}{\nu}$, and $\Lambda$ are all finite, with $a\nu =  \ell (\ell + 1) / m$ and $\Lambda = \frac{\sigma}{\nu} = \ell (\ell + 1)$. Inserting these expressions into the radial equation, Eq.~(\ref{eq:radial}), and multiplying by $f/r^2$, where $f=1-2M/r$, leads to the $s=1$ Regge-Wheeler equation,
 \beq
 f \frac{d}{dr}\left( f \frac{dR}{dr} \right) + \left( \omega^2 - f \frac{\ell (\ell + 1)}{r^2} \right) R = 0 .
 \eeq
This is corroborating evidence that the odd-parity mode is indeed of vector type.

The Schwarzschild limit is examined more closely in Appendix \ref{app:schw}.

\subsubsection{Spectral decomposition method}
With the massless and static limits established, we move on to a method for computing the angular eigenvalue $\nu$ for general $a$ and $\mu$. Following Ref.~\cite{Dolan:2018dqv}, the angular function $S(\theta)$ is decomposed in spherical harmonics $Y^{m}_{\ell'}(\theta)$ as
\beq
S\left(\theta\right)=\sum_{\ell'=0}^{\infty}b_{\ell'}Y_{\ell'}^{m}\left(\theta\right), \quad \quad \ell' = |m| + 2 k' + \eta.
\eeq
The angular equation does not couple harmonics of opposite parity, and so an eigensolution takes a definite parity, and thus is expanded in only either odd or even $\ell$-modes. Here
\beq
\eta = \frac{1}{2} \left( 1 -  (-1)^{\ell + m + P} \right)
\eeq
where $P=0$ for even-parity modes and $P=1$ for odd-parity modes. In other words, $\eta$ takes the value $0$ or $1$, with $\eta = 0$ ($\eta = 1$)  if $\ell + m$ is even (odd) for even parity, and $\eta = 1$ ($\eta = 0$) for odd parity.

This ansatz is substituted into Eq.~(\ref{eq:S2}). The orthogonality of the spherical harmonics is exploited to obtain a matrix equation for the coefficients $b_{k'}$, namely, 
\beq
\sum_{k'=0}^{\infty}M_{kk'}b_{k'}=0 ,
\eeq
where
\beq
M_{kk'}=\left[\Lambda-\ell'\left(\ell'+1\right)\right]\delta_{\ell\ell'}+a^{2}\left[\nu^{2}\ell' \left(\ell' + 1\right)-\nu^{2}\Lambda-2\sigma\nu-q^{2}\right]c_{\ell\ell'}^{(2)}-2a^{2}\nu^{2}d_{\ell\ell'}^{(2)}+q^{2}\nu^{2}a^{4}c_{\ell\ell'}^{(4)} ,
\eeq
$\ell = |m| + 2 k + \eta$ and $\ell' = |m| + 2 k' + \eta$. Here $c_{\ell\ell'}^{(2)}$ and $d_{\ell\ell'}^{(2)}$ are coupling constants, that vanish when $\left|k-k'\right|>1$ and $c_{\ell\ell'}^{(4)}$ is a coupling constant that vanishes when $\left|k-k'\right|>2$. All three constants are defined in Eq.~(32) of Ref.~\cite{Dolan:2018dqv}. In general, the matrix $M_{kk'}$ is pentadiagonal. It becomes tridiagonal in the marginally bound limit $\omega^{2}=\mu^{2}$ and diagonal in the static limit. 
 
The angular eigenvalue $\nu$ is found by seeking the roots of the (truncated) determinant of this matrix using the known massless eigenvalue as the initial guess. 

For the scalar-type polarization, $\nu$ vanishes in the massless limit, for all $\ell$ and $m$, and all five terms in the recurrence relation vanish. As $\nu \rightarrow 0$ irregardless of $\ell$, this leads to a `pile-up' of eigenvalues in the small $\mu$ regime. To handle this issue, we made a change of variables, defining $\tau$ via
\beq
\nu = \frac{\mu^{2}}{\omega}\left(1 + \tau\mu^{2} \right).
\eeq
This is informed by the requirement that, for the scalar-type mode, the angular differential equation (\ref{eq:angular}) should reduce to the $s=0$ spheroidal harmonic equation in the massless limit, which in turn implies that $\tau = -\frac{\lambda_0}{\omega^{2}}$ in this limit, where $\lambda_0$ is the Teukolsky $s = 0$ eigenvalue. We can then repeat the same procedure used to calculate $\nu$ in the vector polarizations to calculate $\tau$, with $-\frac{\lambda_0}{\omega^{2}}$ as an initial guess.

%
%
%

\subsection{Numerical method\label{sec:numerical}}

\subsubsection{Calculating QNM frequencies}
Naively, the series coefficients $a_k$ in Eq.~(\ref{eq:ansatz}) can be found by solving the recurrence relations in Eq.~(\ref{eq:5term}) iteratively, starting with $\alpha_{-1} a_1 + \beta_{-1} a_0 = 0$ and $a_0 = 1$ at the first step. However, for general $\omega$, the series $\sum^\infty_k a_k$ is divergent, and the outgoing boundary condition is not satisfied. For a QNM frequency $\omega^{(Q)}$, the series is convergent in principle, but divergent in practice under forward recursion due to the accumulation of numerical error.

A robust procedure for calculating QNM frequencies from $n$-term recurrence relations was presented by Leaver \cite{Leaver:1985ax,Leaver:1990zz}. The first step is to apply Gaussian elimination \cite{Leaver:1990zz} to the $n$-term relation to reduce it to a 3-term relation of the form
\begin{align}
\tilde\alpha_{0} a_1 + \tilde\beta_{0} a_0 &= 0 , \nn \\
\tilde\alpha_{n} a_{n+1} + \tilde\beta_{n} a_{n} + \tilde\gamma_n a_{n-1} &= 0 . \label{eq:3term}
\end{align}
This step is described in more detail in Appendix \ref{appendix}. The second step is to seek a solution sequence to the three-term recurrence relation that is minimal as $n \rightarrow \infty$. This is equivalent \cite{Leaver:1985ax} to seeking solutions of the continued-fraction equation
\beq
0 = \tilde{\beta}_0 - \frac{\alpha_0 \gamma_1}{\tilde{\beta}_1 - }\frac{\tilde\alpha_1 \tilde\gamma_2}{\tilde\beta_2 - }\frac{\tilde\alpha_2 \tilde\gamma_3}{\tilde\beta_3 - } \ldots \label{eq:ctdfrac}
\eeq
or one of its inversions \cite{Leaver:1985ax}. Typically, the $n$th quasi-normal mode is the most numerically stable root of the $n$th inversion \cite{Leaver:1985ax}. The continued fraction is evaluated to the desired precision using the modified Lentz algorithm \cite{Press:1988}. Numerical solutions for $\omega$ such that (\ref{eq:ctdfrac}) is satisfied may be found with standard root-finding algorithms. As an initial value for the algorithm we typically used the $s=1$ (vector) and $s=0$ (scalar) massless QNM frequencies \cite{Berti:ringdown}.

As a consistency check, we also evaluated the series coefficients $a_n$ by applying forward recurrence directly to the 5-term relation (\ref{eq:5term}). At a QNM frequency, the series coefficients $a_n$ typically decrease ($|a_{n+1}| < |a_n|$) up to some large value of $n$; but beyond this point accumulated numerical error leads to renewed growth in the series coefficients. 
We confirmed that the QNM frequencies found via the continued fraction method are close to the minima in a merit function $\log |a_n|$ for sufficiently large $n$. 

 \subsubsection{Validation}

We made two consistency checks on the integrity of our numerical code and recurrence relation. 
First, we computed the spectrum of the even-parity QNMs in the Schwarzschild case ($a = 0$), and compared with the data sets in Ref.~\cite{Rosa:2011my}, finding agreement to 9 significant figures (a direct comparison of the odd-parity QNMs was not possible, due to the divergence of $\nu$ in the static limit). Second, we computed quasi-bound state frequencies of the Proca field on the Kerr spacetime, using the method above with the opposite sign choice for $q$ in Eq.~(\ref{eq:qdef}). We found agreement with the results of Ref.~\cite{Dolan:2018dqv} to at minimum 6 significant figures.

\section{Results\label{sec:results}}

Here we present a selection of numerical results for the QNM spectrum of the Proca field on Kerr spacetime. 

\begin{figure}
	\includegraphics[width=14cm]{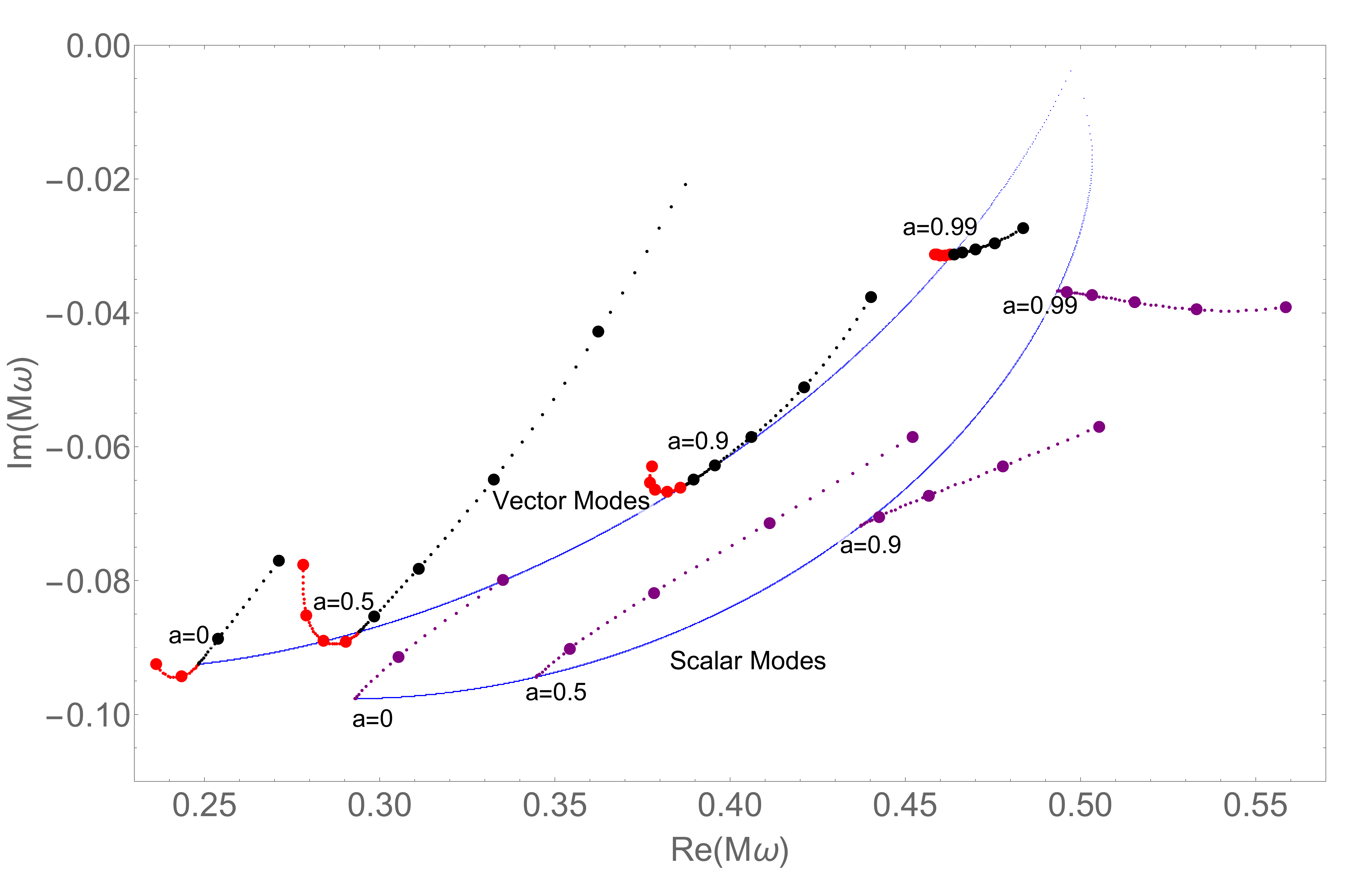}
	\caption{Fundamental QNMs of massless and massive vector fields in the complex plane,
		$\ell = m = 1, n=0$. The blue curves show the QNMs of the massless vector (electromagnetic)
		and scalar fields for varying $a$. On the upper curve, the black points show the odd-parity Proca QNMs, and the red points show the even-parity Proca QNMs of vector type. On the lower curve, the purple points show QNMs of even-parity scalar type. The mass spacing between large (small) points is $M\mu = 0.1$ ($0.01$). 
		}
\label{fig:l1n0vector}
\end{figure}

Figure \ref{fig:l1n0vector} shows the fundamental $\ell = m = 1$ QNM frequencies of the Proca field in all three polarizations, for a range of black hole spin parameters $a$. In the massless limit, the QNM frequencies of the two vector-type polarizations coincide with the QNM frequency of the electromagnetic field (upper
blue curve in Fig.~\ref{fig:l1n0vector}) and the QNM frequency of the scalar-type polarization coincides with the frequency of the massless scalar field (lower blue curve in Fig.~\ref{fig:l1n0vector}). The general trend for the odd-parity and scalar-type modes is for the oscillation frequency $\text{Re} (\omega)$ to increase, and for the damping rate $-\text{Im}(\omega)$ to decrease, with increasing field mass $\mu$ and with increasing black hole spin $a$. This is broadly the same trend as seen for the massive scalar field in Fig.~\ref{fig:s0QNMkerr}, and may be understood with reference to Eq.~(\ref{eq:WKB}) and Fig.~\ref{fig:Vplot}. Bucking this trend, the even-parity vector mode decreases in frequency and increases in damping rate for small $\mu$. It is also notable that, at high spin (e.g.~$a=0.99M$), the damping rate of the even-parity scalar mode increases with $\mu$ for small $\mu$.

As the field mass increases, the QNMs become evanescent (see also Fig.~\ref{fig:Vplot} and \ref{fig:s0QNMschw}). The QNM frequencies move towards the real axis, and thus towards \emph{quasiresonance}. However, as noted in Ref.~\cite{Decanini:2016ifm},  evanescent modes will play an insignificant role in the response of the black hole to an initial perturbation, in comparison to the low-lying bound states and the propagative QNMs in higher multipoles.


\begin{figure}
	\includegraphics[width=14cm]{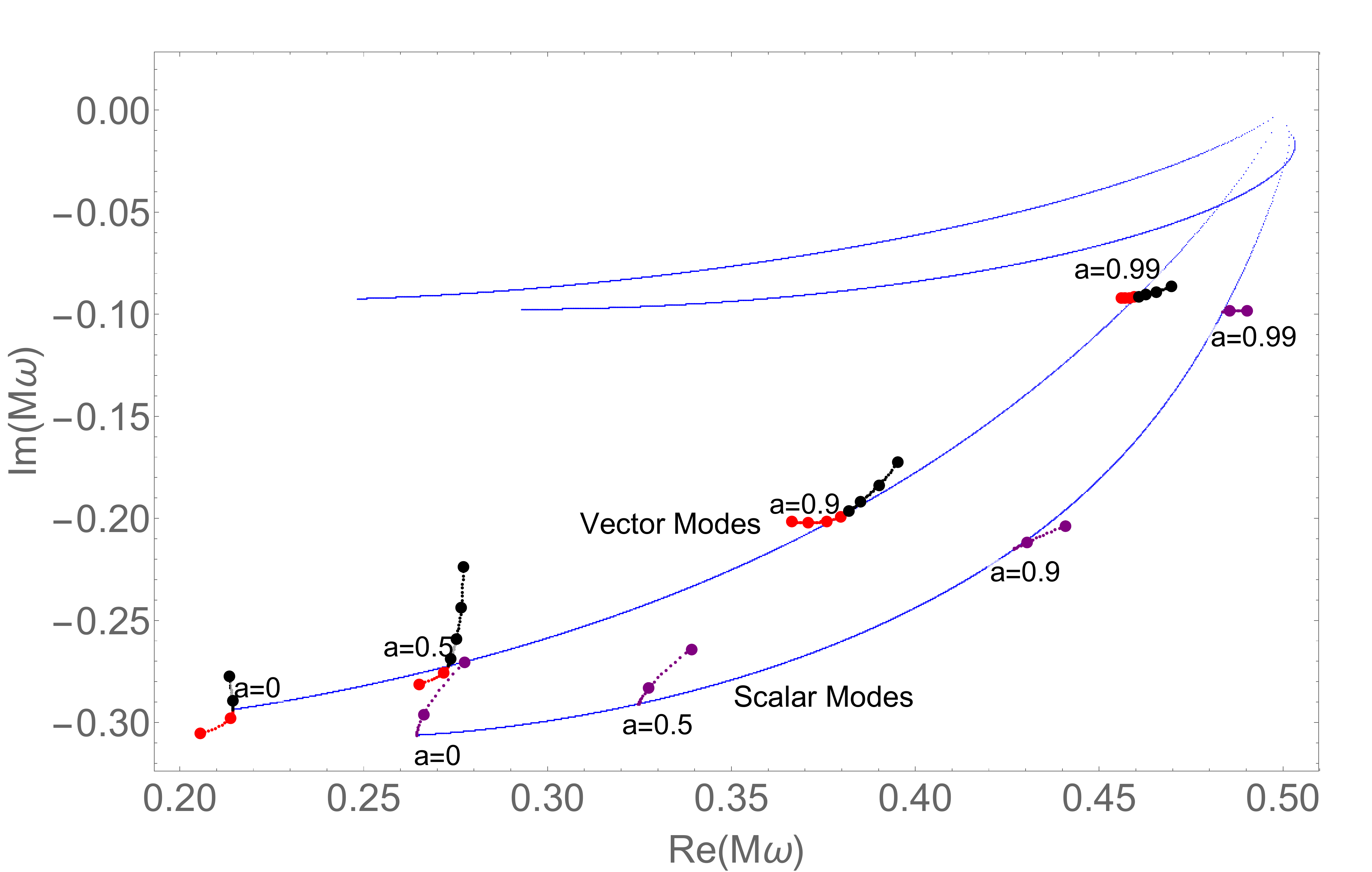}
	\caption{As for Fig.~\ref{fig:l1n0vector} but for the first overtone $n=1$. The thin blue lines are the massless QNMs of the fundamental mode shown in Fig.~\ref{fig:l1n0vector}.}
\label{fig:l1n1vector}
\end{figure}

Figure \ref{fig:l1n1vector} shows QNM frequencies for the first overtone ($n=1$) of the $\ell = m = 1$ mode on Kerr. As for the fundamental mode, the introduction of a field mass leads to a migration towards higher frequencies and lower dampings for the odd parity and scalar-type modes, with the opposite trend in evidence for the even-parity vector mode.

\begin{figure}
	\includegraphics[width=14cm]{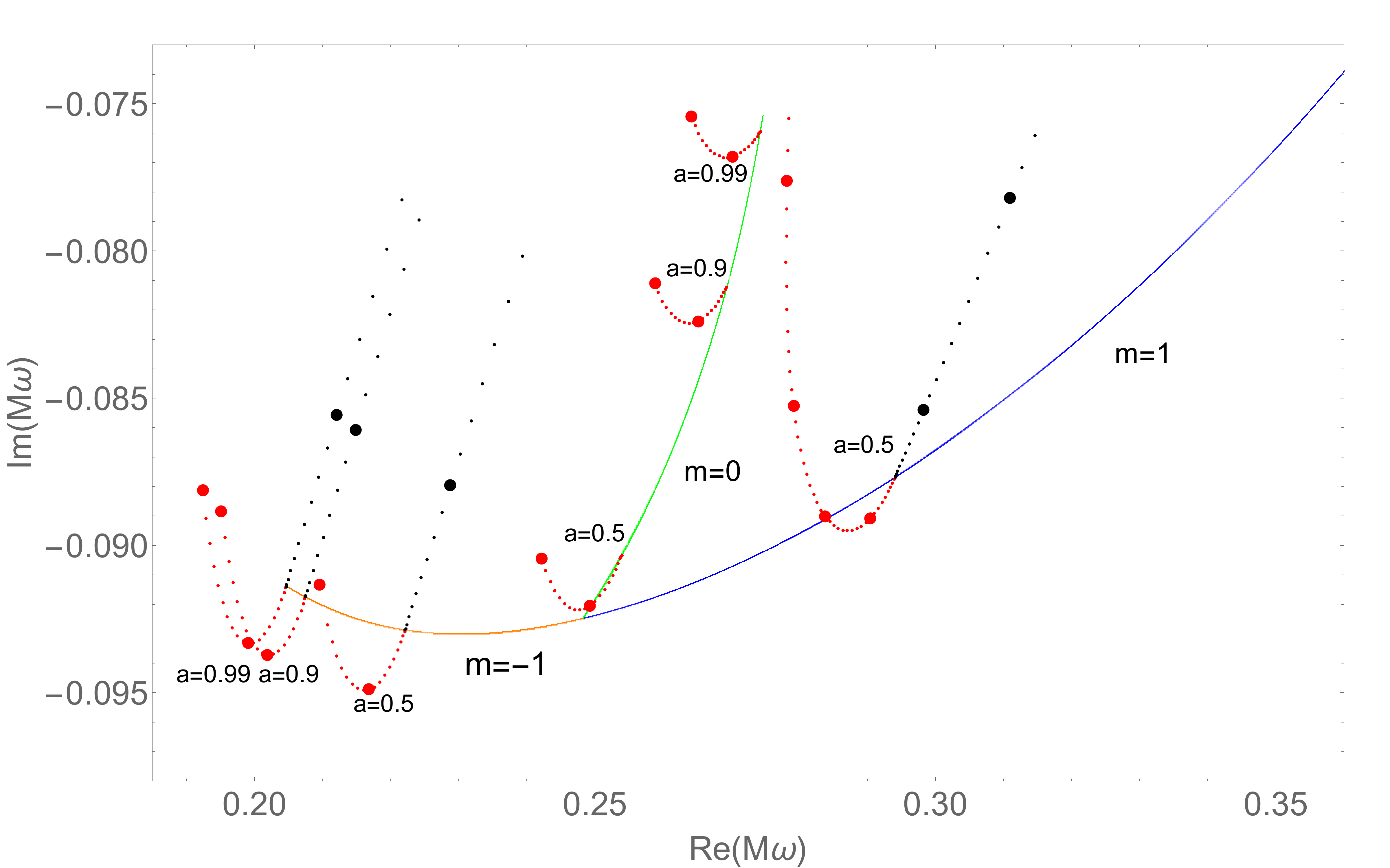}\caption{Vector-type QNM frequencies for $m = -1$ (left, orange), $m=0$ (centre, green) and $m=1$ (right, blue) dipole ($\ell = 1$) modes, with a mass spacing $\Delta(M\mu) = 0.01$. The plot shows the detail of the $m=-1$ and $m=0$ cases; the $m=1$ cases for higher $a$ are shown in Fig.~\ref{fig:l1n0vector}. 
	}
	\label{fig:vectorl1}
	\end{figure}

\begin{figure}
	\includegraphics[width=14cm]{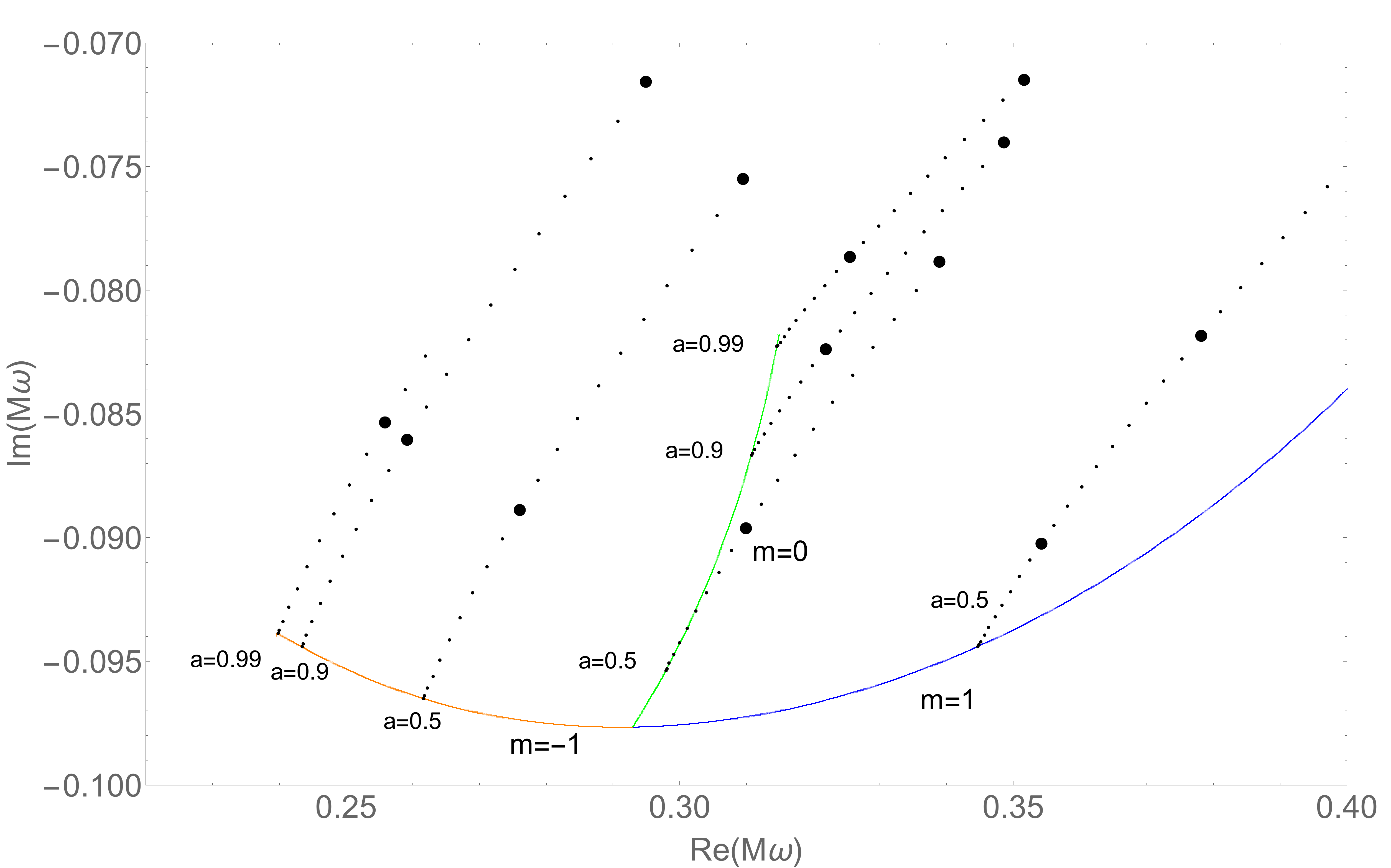}\caption{The scalar-type Proca QNMs for the $m=-1$, $m=0$ and $m=1$ branches of the $\ell = 1$, $n=0$ spectrum. As in Fig.~\ref{fig:l1n0vector}, the mass spacing between large (small) points is $M\mu = 0.1$ ($0.01$). 
	}
	\label{fig:scalarl1}
\end{figure}

Figues \ref{fig:vectorl1} and \ref{fig:scalarl1} show the fundamental dipole frequencies for $m=1$, $m=0$ and $m=-1$, for the vector-type and scalar-type QNMs, respectively. The $m=-1$ modes exhibit lower oscillation frequencies than the $m=1$ modes, as also seen in the scalar-field case in Fig.~\ref{fig:s0QNMkerr}. In the semi-classical picture, $m < 0$ modes are associated with geodesic orbits that pass around the black hole in the opposite sense to its rotation and such orbits have lower orbital frequencies than their co-rotating counterparts.

\begin{figure}
 \includegraphics[width=14cm]{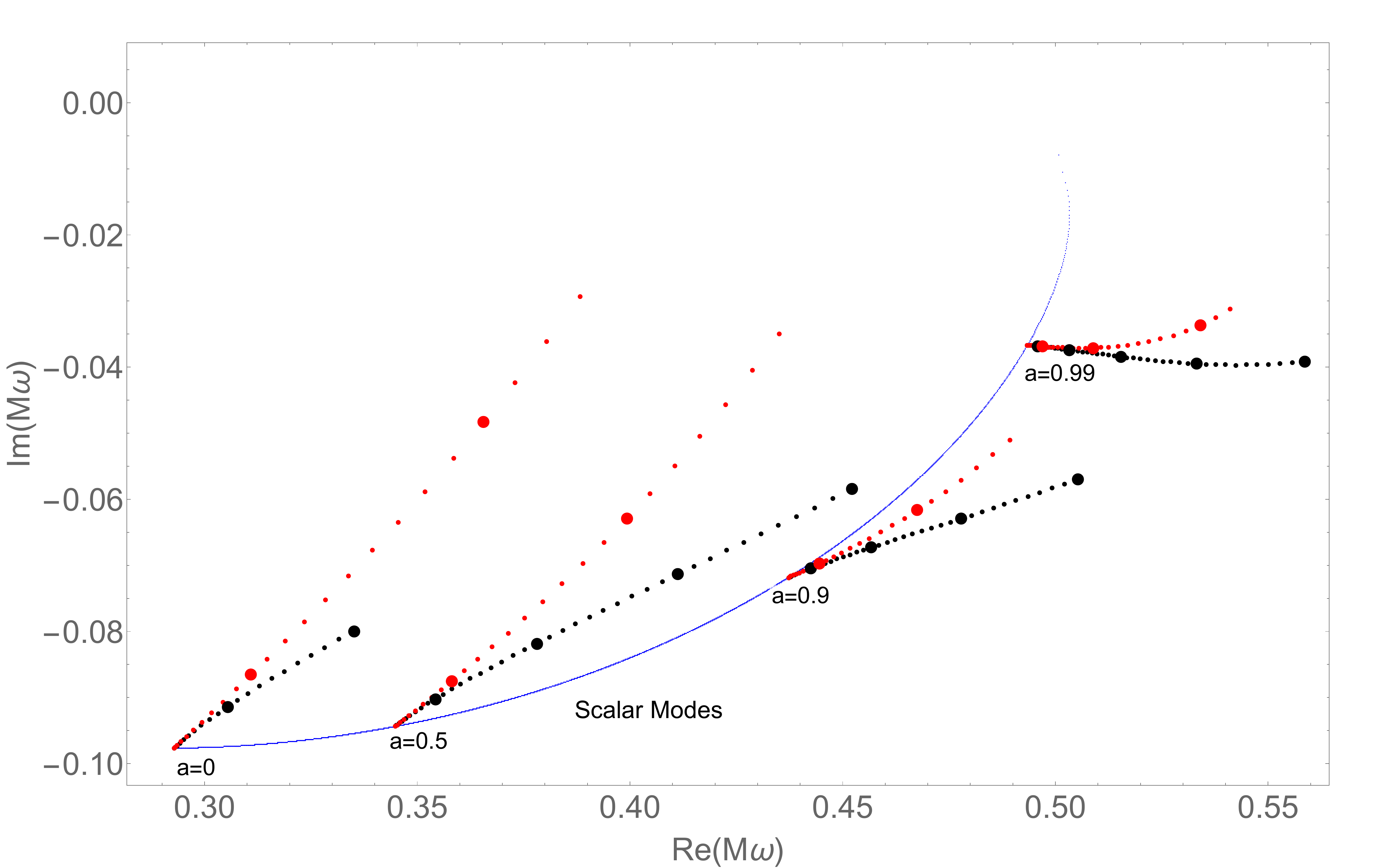}
 \caption{Comparing the QNM spectrum of the scalar-type polarization of the Proca field (black) with the QNM spectrum of the massive scalar field (red). }
 \label{fig:scalar-comparison}
\end{figure}

Figure \ref{fig:scalar-comparison} shows a comparison between the spectrum of the scalar-type polarization of the Proca field and the spectrum of a massive scalar field. We observe that, for small masses, the trajectories of the QNMs in the complex plane are closely aligned. At higher masses, the branches diverge from one another and the Proca modes typically show higher oscillation frequencies and faster damping rates than the scalar-field modes. 

\begin{figure}
    \includegraphics[width=14cm]{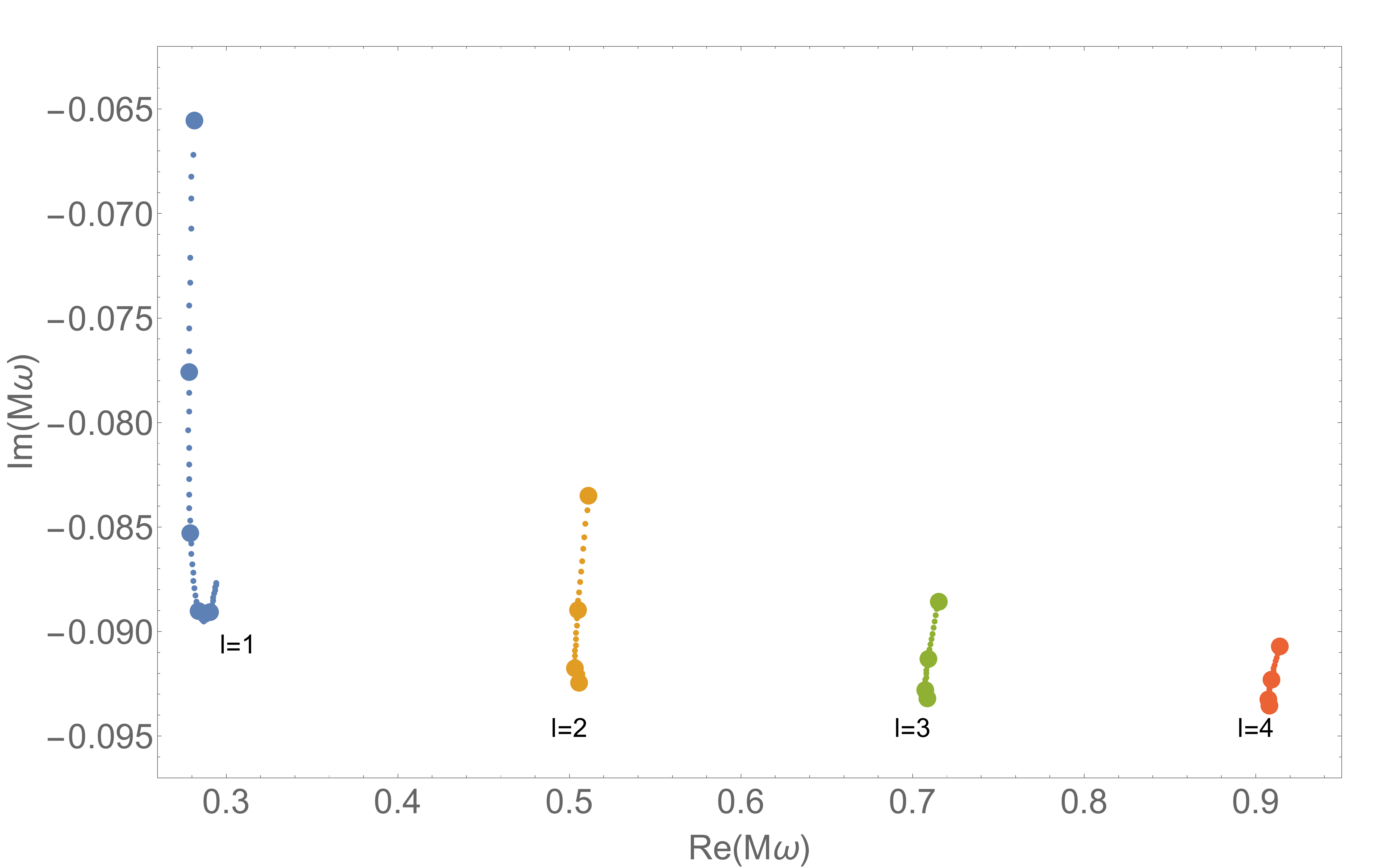}
    \caption{The higher multipoles of the even-parity vector QNMs for $a=0.5M$ and $m=1$.}
    \label{fig:higher-multipoles}
\end{figure}

Figure \ref{fig:higher-multipoles} shows some examples of the fundamental ($n=0$) QNM frequencies of higher multipoles ($\ell = 1$, $2$, $\ldots$). For small masses $M\mu$, the even-parity vector mode actually increases in damping rate and decreases in frequency. However, for larger masses the damping rate decreases, and each branch migrates towards the real axis; similar behaviour is shown in Fig.~\ref{fig:s0QNMschw} for the scalar-field Schwarzschild case.

Table \ref{tbl} contains some sample QNM frequencies and their respective angular eigenvalues. 

\begin{table}
	\begin{tabular}{|c|c|c|c|c|c|c|}
		\hline 
		polarization & $l$ & $n$ & $\Re\left(M \omega\right)$ & $\Im\left(M \omega\right)$ & $\Re\left(M \nu\right)$ & $\Im\left(M \nu\right)$\tabularnewline
		\hline 
		\hline 
		even, vector & 1 &  0 & 0.290441 & -0.0890795 & -0.201844 & 0.0587629\tabularnewline
		\hline 
		even, vector & 1 &  1 & 0.271840 & -0.275480 & -0.148270 & 0.174811\tabularnewline
		\hline 
		odd, vector & 1 & 0 & 0.298286 & -0.0853954 & 3.85422 & 0.0407235\tabularnewline
		\hline 
		odd, vector & 1 &  1 & 0.273674 & -0.269075 & 3.86338 & 0.128896\tabularnewline
		\hline 
		scalar & 1 &  0 & 0.354243 & -0.0902282 & 0.0243336 & 0.00493117\tabularnewline
		\hline 
		scalar & 1 &  1 & 0.327587 & -0.283202 & 0.0185236 & 0.0133421\tabularnewline
		\hline 
	\end{tabular}\caption{Sample quasinormal mode frequencies and angular eigenvalues for the parameters $a=0.5$, $\mu=0.1$,
		$\ell =1$, $m=1$. }
\label{tbl}
\end{table}


\section{Discussion and conclusions\label{sec:conclusions}}

In the preceding sections we have computed the low-lying QNM frequencies of the (neutral) Proca field on the Kerr spacetime, for the first time. We find that the degeneracy of the two vector modes of the electromagnetic QNM spectrum is split by the introduction of a field mass, and a third scalar (longitudinal) polarization state arises. We have shown how the QNM frequencies migrate in the complex plane as the field mass is increased, in a somewhat similar manner to the modes of the massive scalar field, but with subtleties associated with spin and polarization. As in the scalar-field case, there is a transition from propagative to evanescent behaviour as the mass increases.

The achievements herein are primarily technical, extending the calculation of QNMs to a new domain that combines field mass, spin and the frame-dragging of spacetime. The calculation was made possible by the complete separation of variables achieved by Frolov, Krtou{\v{s}} and Kubiz{\v{n}}\'ak for the Proca field on Kerr spacetime \cite{Frolov:2018eza,Frolov:2018ezx}. This reduced the problem to that of imposing boundary conditions on a pair of second-order ordinary differential equations. We have shown here that, with a suitable ansatz, the problem of finding QNMs reduces to the problem of finding convergent solutions to a five-term recurrence relation (\ref{eq:5term}); and this can be handled with the standard methods of Gaussian elimination and the evaluation of a continued-fraction via the modified Lentz algorithm. The numerical results appear robust and accurate.

The five-term recurrence relation (\ref{eq:5term}) also yields the quasi-bound state spectrum recently studied in Refs.~\cite{Frolov:2018ezx, Dolan:2018dqv, Siemonsen:2019ebd, Baumann:2019eav, Witek:2012tr, Pani:2012bp, Pani:2012vp, Baryakhtar:2017ngi,East:2017mrj, East:2017ovw, Cardoso:2018tly}. One may employ the numerical method exactly as presented here, but with the opposite choice of sign in Eq.~(\ref{eq:qdef}). Our method is more accurate and robust than the direct integration method used in Ref.~\cite{Dolan:2018dqv}, and is complementary to the spectral method employed in Ref.~\cite{Baumann:2019eav}.  

The prospect of observing the Proca QNM spectra in nature seems remote, not least because of the apparent absence of spin-one fields with sufficiently small mass. For an astrophysical black hole, $M\mu$ is exceedingly large for vector bosons in the Standard Model, and exceedingly small or zero for the photon, as can be seen by reinstating dimensionful constants:
\beq
M \mu = \frac{M \mu}{m_P^2} \approx 7.52 \cdot 10^{9} \times \left( \frac{M}{M_{\odot} } \right) \left( \frac{\mu c^2}{\text{eV}} \right) .
\eeq
For a black hole of mass $10M_\odot$ and a W-boson, one has $M\mu \approx 6 \times 10^{21}$, 
conversely, for a BH of same mass and a massive photon, one has $M\mu \lesssim 2 \times 10^{-16}$ (assuming a photon mass $\lesssim 3 \times 10^{-27} \text{eV}/\text{c}^{2}$). 
In the former case, all modes with $\ell + 1/2 \lesssim O(M\mu)$ will be evanescent. In the latter case, the QNM spectrum will in effect be identical to the spectrum of the electromagnetic field, but with one key difference: an additional longitudinal polarization with the QNM spectrum of a scalar field, if $\mu > 0$. 

There are a variety of mechanisms by which a field can acquire an effective mass. For example, in the presence of a strong magnetic field \cite{Konoplya:2007yy}, in Horndeski gravity and other extensions of General Relativity \cite{Tattersall:2018nve}, and in string theory compactifications and theories with ``large'' extra dimensions. Ultra-light fields with masses $\mu \ll \text{eV}/c^{2}$ are often considered as plausible dark-matter candidates \cite{Hui:2016ltb}. A well-known example is the (hypothetical) axion, a pseudoscalar introduced to solve the strong CP problem of QCD. Axion-like particles with masses that are \emph{not} linked to the axion decay constant emerge from string-theory-inspired theories, with compactification mechanisms that generate a landscape of ultralight axions, known as the ``string axiverse'' \cite{Arvanitaki:2009fg}, on mass scales possibly down to the present Hubble scale. Massive hidden $U(1)$ vector fields are also a generic feature of BSM scenarios, again particularly from string theory compactifications. For stellar-size black holes ($M = 5$--$20M_{\odot}$) and ultralight boson(s) in the range $\mu = 10^{-9}$--$10^{-13} \, \text{eV}/c^2$, the fundamental propagative QNMs would be significantly altered by the field mass. However, the existence of a boson in this mass range would also lead to the inflation of `boson clouds', triggered by the exponential growth of quasibound states in the superradiant regime \cite{Pani:2012bp,Pani:2012vp}. The latter is the dominant phenomenon, and the priority for those seeking experimental signatures of ultralight bosons.

Finally, two extensions of this work suggest themselves. First, an exploration of the QNM spectrum of the Proca field on a charged, rotating black hole spacetime, i.e., the Kerr-Newman solution. In that case, a slightly more general five-term relation emerges, and there is a larger parameter space to explore. Again, the method could also be used to compute the quasi-bound states, complementing the method of Ref.~\cite{Cayuso:2019ieu}. Second, an investigation of the Proca spectrum in the approach to extremality ($a \rightarrow M$), where branching of QNMs has been found in the electromagnetic case \cite{Yang:2012pj,Yang:2013uba}.

 \appendix

\section{Gaussian elimination of 5-term recurrence relation\label{appendix}}

The recurrence relation can be written in matrix form, $\mathbf{M} \mathbf{a} = 0$, as follows:
\begin{align}
\begin{pmatrix}
\beta_{-1} & \alpha_{-1} & \cdot & \cdot & \cdot & \cdot & \ldots \\
\gamma_{0} & \beta_{0} & \alpha_{0} & \cdot & \cdot & \cdot  & \ldots \\
\delta_{1} & \gamma_{1} & \beta_{1} & \alpha_{1} & \cdot & \cdot & \ldots  \\
\epsilon_{2} & \delta_{2} & \gamma_{2} & \beta_{2} & \alpha_{2} & \cdot & \ldots  \\
\cdot & \epsilon_{3} & \delta_{3} & \gamma_{3} & \beta_{3} & \alpha_{3} & \ldots \\
\ldots & \ldots & \ldots & \ldots & \ldots & \ldots & \ldots 
\end{pmatrix}
\begin{pmatrix} a_0 \\ a_1 \\ a_2 \\ a_3 \\ a_4 \\ \ldots \end{pmatrix}
=
\begin{pmatrix} 0 \\ 0 \\ 0 \\ 0 \\ 0 \\ \ldots \end{pmatrix}.
\end{align}
A quasinormal mode corresponds to a frequency such that $\text{det} \, \mathbf{M} = 0$. We now perform row operations on this system of equations. The first step is to eliminate $\epsilon_n$, using
\beq
\epsilon'_k = 0, \quad 
\delta'_k = \delta_k - \frac{\epsilon_k \gamma'_{k-1}}{\delta'_{k-1}} , \quad
\gamma'_k = \gamma_k - \frac{\epsilon_k \beta'_{k-1}}{\delta'_{k-1}} , \quad
\beta'_k = \beta_k - \frac{\epsilon_k \alpha'_{k-1}}{\delta'_{k-1}} , \quad
\alpha'_k = \alpha_k , 
\eeq
for $k \ge 2$ (and $\delta'_k = \delta_k$, etc., for $k < 2$). The next step is to eliminate $\delta^\prime_n$ using
\beq
\epsilon''_k = \delta''_k = 0, \quad 
\gamma''_k = \gamma'_k - \frac{\delta'_k \beta''_{k-1}}{\gamma''_{k-1}} , \quad
\beta''_k = \beta'_k - \frac{\delta'_k \alpha''_{k-1}}{\gamma''_{k-1}} , \quad
\alpha''_k = \alpha'_k , 
\eeq
for $k \ge 1$ (and $\gamma'_k = \gamma_k$, etc., for $k < 1$). This leaves the determinant of the matrix $\mathbf{M}$ in the form
\begin{align}
\begin{vmatrix}
\beta_{-1} & \alpha_{-1} & \cdot & \cdot & \cdot & \cdot & \ldots \\
\gamma_{0} & \beta_{0} & \alpha_{0} & \cdot & \cdot & \cdot  & \ldots \\
\cdot & \gamma''_{1} & \beta''_{1} & \alpha_{1} & \cdot & \cdot & \ldots  \\
\cdot & \cdot & \gamma''_{2} & \beta''_{2} & \alpha_{2} & \cdot & \ldots  \\
\cdot & \cdot & \cdot & \gamma''_{3} & \beta''_{3} & \alpha_{3} & \ldots \\
\ldots & \ldots & \ldots & \ldots & \ldots & \ldots & \ldots 
\end{vmatrix}
\end{align}
using here that $\alpha''_k = \alpha_k$. As this matrix is now tridiagonal, the system of equations represents a three-term relation and the continued-fraction method can be applied. The coefficients in Eq.~(\ref{eq:3term}) are given by $\tilde{\alpha}_{k} = \alpha''_{k-1}$, and likewise for $\beta$ and $\gamma$.

 \section{The Schwarzschild limit\label{app:schw}}
In this section we link the separation ansatz (\ref{eq:ansatz}) in the Schwarzschild limit $(a \rightarrow 0)$ to the earlier approach of Rosa \& Dolan. In Ref.~\cite{Rosa:2011my}, a separation of variables for the Schwarzschild case was performed using the vector spherical harmonics
\begin{eqnarray} \label{vector-harmonics}
Z_{\mu}^{(1)\ell m} &=& \left[ 1, 0, 0, 0 \right] Y \\
Z_{\mu}^{(2)\ell m} &=& \left[ 0, f^{-1}, 0, 0 \right]] Y \\
Z_{\mu}^{(3)\ell m} &=& {r\over\sqrt{\ell(\ell+1)}} \left[ 0, 0, \partial_\theta, \partial_\phi  \right] Y \\
Z_{\mu}^{(4)\ell m} &=&  {r\over\sqrt{\ell(\ell+1)}} \left[ 0, 0, \frac{1}{\sin \theta} \partial_\phi, - \sin \theta \partial_\theta \right] Y,   
\end{eqnarray}
where $Y \equiv Y_{\ell m}(\theta, \phi)$ are scalar spherical harmonics. Decomposing the vector potential in this basis,
\beq
A_{\mu}(t,r,\theta,\phi) = \frac{1}{r} \sum_{i=1}^{4} \sum_{\ell m} c_i \, u^{\ell m}_{(i)}(t,r) Z_\mu^{(i) \ell m}(\theta, \phi) \, ,
\label{sep-ansatz}
\eeq
where $c_1 = c_2 = 1$, $c_3 = c_4 = [\ell (\ell+1)]^{-1/2}$ leads to a set of four second-order partial differential equations,
\begin{subequations}
\begin{align}
\Op u_{(1)} &+ \left[ \frac{2M}{r^2} \left(  \dot{u}_{(2)} - u^\prime_{(1)} \right) \right] = 0 , \\
\Op u_{(2)} &+ \frac{2}{r^2}\left[ \left( M \dot{u}_{(1)} - u^\prime_{(2)} \right) - f^2\left( u_{(2)} - u_{(3)} \right) \right]=0 , \\
\Op u_{(3)} &+ \left[ \frac{2 f \ell (\ell+1)}{r^2} u_{(2)} \right] = 0 ,  \\
\Op u_{(4)} &= 0~, 
\end{align}
\label{eq:uode}
\end{subequations}
along with the first-order Lorenz condition,
\beq
-\dot{u}_{(1)} + u^\prime_{(2)} + \frac{f}{r} \left(u_{(2)} - u_{(3)}  \right)  =  0 ,   \label{Lorenz-eq}
\eeq
where $f = 1-2M/r$ and
\beq
\Op \equiv -\frac{\partial^2}{\partial t^2} + \frac{\partial^2}{\partial r_\ast^2} - f \left[ \frac{\ell(\ell+1)}{r^2} + \mu^2 \right]~, 
\eeq
and $\dot u \equiv \tfrac{\partial u}{\partial t} = -i \omega u$, $u^\prime \equiv \tfrac{\partial u}{\partial r_\ast}$ and the tortoise coordinate $r_\ast$ is defined via $dr_\ast = f^{-1} dr$.

\subsection{Even parity}
A straightforward comparison of Eq.~(\ref{eq:ansatz})--(\ref{eq:Babalt}) and Eq.~(\ref{sep-ansatz}) shows that
\begin{subequations}
\begin{align}
u_1(r) &= - \frac{i f r \left( \nu r \partial_r + \omega / f \right) R}{\mathfrak{q}_r}, \\
u_2(r) &= \frac{f r \left(\partial_r - \omega \nu r / f \right) R}{\mathfrak{q}_r} , \\
u_3(r) &= \Lambda R , \\
u_4(r) &= 0,
\end{align}
\label{eq:ueven}
\end{subequations}
and $S(\theta) = Y_{\ell m}$, $\Lambda \equiv \mu^2 / \nu^2 - \omega / \nu = \ell (\ell + 1)$ and $\nu$ is given in Eq.~(\ref{eq:nustatic}).
Consistency was checked by substituting (\ref{eq:ueven}) into (\ref{eq:uode})--(\ref{Lorenz-eq}) and employing (\ref{eq:radial}).

\subsection{Odd parity}
Additional care is required for the odd-parity sector since, as noted in Sec.~\ref{sec:polarization}, the angular eigenvalue $\nu$ diverges as $a \rightarrow 0$, but $m a \nu$ approaches a constant. A direct comparison of Eq.~(\ref{eq:ansatz})--(\ref{eq:Babalt}) and Eq.~(\ref{sep-ansatz}) yields $u_1=u_2=u_3=0$, $u_4(r) = R(r)$ and two equations for the angular function,
\begin{align}
\left( \sin \theta \partial_\theta + m a\nu \cos \theta \right) S &= \frac{i m \mathfrak{q}_\theta}{\ell (\ell + 1)} Y_{\ell m} , \\
\left( m + a \nu \sin \theta \cos \theta \partial_\theta \right) S &= \frac{i \sin \theta \, \mathfrak{q}_\theta}{\ell (\ell + 1)} \partial_\theta Y_{\ell m} . 
\end{align}
These are consistent with Eq.~(\ref{eq:angular}) in the $a \rightarrow 0$ limit if and only if $a \nu = \ell (\ell + 1) / m$.
A consistent solution is
\beq
S(\theta) = \frac{i}{ \ell (\ell + 1) m } \left( \sin \theta \, \partial_\theta -  \ell (\ell + 1) \cos \theta \right) Y_{\ell m} . 
\eeq
Employing the properties of associated Legendre polynomials, we establish that
\beq
S(\theta) \propto \ell^2 (\ell + 1 - m) P_{\ell + 1}^m(\cos \theta) + (\ell + 1)^2(\ell + m) P_{l-1}^m(\cos \theta) ,
\eeq
which makes it clear that $S(\theta)$ is odd parity.

\acknowledgments
With thanks to Mohamed Ould Elhadj and Tom Stratton for advice on numerical methods and implementation. 
S.R.D.~acknowledges financial support from the European Union's Horizon 2020 research and innovation programme under the H2020-MSCA-RISE-2017 Grant No.~FunFiCO-777740, and from the Science and Technology Facilities Council (STFC) under Grant No.~ST/P000800/1.

\bibliography{refs}
\bibliographystyle{apsrev4-1}

\end{document}